\documentclass[twocolumn,english,floatfix,amsmath,amssymb,reprint,nofootinbib,prb,superscriptaddress]{revtex4-1}

\usepackage[]{fontenc}
\usepackage[latin9]{inputenc}
\usepackage{color}
\usepackage{babel}
\usepackage{amsmath}
\usepackage{amssymb}
\usepackage{graphicx}
\usepackage{wasysym}
\usepackage[unicode=true,pdfusetitle,
 bookmarks=true,bookmarksnumbered=false,bookmarksopen=false,
 breaklinks=true,pdfborder={0 0 0},backref=false,colorlinks=true]
 {hyperref}
\hypersetup{
 linkcolor=blue,citecolor=blue,urlcolor=blue}
\usepackage{breakurl}

\makeatletter

\newcommand*\LyXZeroWidthSpace{\hspace{0pt}}

\usepackage{babel}
\usepackage{amscd}
\usepackage{bm}
\usepackage{graphics}
\usepackage{psfrag}\usepackage{psfrag}\usepackage{lipsum}

\usepackage{pslatex}
\usepackage{color}
\usepackage{babel}

\makeatother

\begin{document}

\title{Adaptive Density Matrix Renormalization Group for Disordered Systems}

\author{J. C. Xavier }

\affiliation{Universidade Federal de Uberlândia, Instituto de Física, C. P 593,
38400-902 Uberlândia, MG, Brazil }

\author{José A. Hoyos }

\affiliation{Instituto de Física de São Carlos, Universidade de São Paulo, C.
P. 369, 13560-970 São Carlos, SP, Brazil}

\author{E. Miranda }

\affiliation{Instituto de Física Gleb Wataghin, Unicamp, Rua Sérgio Buarque de
Holanda, 777, CEP 13083-970 Campinas, SP, Brazil}

\date{\today{}}
\begin{abstract}
We propose a simple modification of the density matrix renormalization
group (DMRG) method in order to tackle strongly disordered quantum
spin chains. Our proposal, akin to the idea of the adaptive time-dependent
DMRG, enables us to reach larger system sizes in the strong disorder
limit by avoiding most of the metastable configurations which hinder
the performance of the standard DMRG method. We benchmark our adaptive
method by revisiting the random antiferromagnetic XXZ spin-1/2 chain
for which we compute the random-singlet ground-state average spin-spin
correlation functions and von Neumann entanglement entropy. We then
apply our method to the bilinear-biquadratic random antiferromagnetic
spin-1 chain tuned to the antiferromagnet and gapless highly symmetric
SU(3) point. We find the new result that the mean correlation function
decays algebraically with the same universal exponent $\phi=2$ as
the spin-1/2 chain. We then perform numerical and analytical strong-disorder
renormalization-group calculations, which confirm this finding and
generalize it for any highly symmetric SU($N$) random-singlet state. 
\end{abstract}
\maketitle

\section{INTRODUCTION}

The theoretical investigation of strongly correlated systems by unbiased
(i.e., whose error is controlled) methods is a challenge, mainly,
due to the lack of appropriate techniques to study those systems.
In the last years some progress has been obtained by methods based
on tensor network states (TNT), such as the multiscale entanglement
renormalization ansatz (MERA)~\citep{Vidalmera} and projected entangled
pair states (PEPS),~\citep{verstraeteadvphys} and by Monte-Carlo-based
methods (for a review, see Refs.~\onlinecite{newman-MC,sandvik-QMC}).

In the case of one dimensional systems, the density matrix renormalization
group (DMRG)~\citep{white} is a remarkable technique capable of
providing quasi-exact results for both static and dynamic properties.~\citep{reviewdmrgS}
(Quantum Monte Carlo techniques are also powerful in $d=1$, but are
limited to some classes of problems due to the famous ``sign problem''.)
In particular, the rich low-energy physics of several ``clean''
systems, belonging to the Tomonaga-Luttinger liquid universality class,~\citep{voit}
was shown to be captured by the DMRG technique.~\citep{reviewdmrgS}

The effects of inhomogeneities, common in real materials, add to the
plethora of phenomena in strongly interacting systems. They can completely
change the critical behavior and induce Griffiths phases surrounding
critical points (for a review, see Refs.~\onlinecite{miranda-dobrosavljevic-rpp05,vojta-review06}).
Among all the exotic phenomena induced by disorder in strongly correlated
systems, one is of particular importance: the infinite-randomness
criticality. In the renormalization-group sense, the concept of infinite-randomness
criticality states that the effective disorder strength of a system
(measured by some statistical fluctuations of a local quantity) increases
without bounds as the systems is probed (coarse grained) on ever larger
length scales. Along the years, it was shown that this concept is
more ubiquitous than previously thought, ranging from spin chains,~\citep{fisher92,fisher94-xxz}
higher dimensional magnetic and superconducting systems,~\citep{motrunich-ising2d,hoyos-kotabage-vojta-prl07}
to non-equilibrium~\citep{hooyberghs-prl,vojta-hoyos-epl15} and
driven systems.~\cite{monthus-floquet,berdanier-etal-pnas18} Interestingly,
there is one biased (approximate) technique capable of studying this
phenomenon: the strong-disorder renormalization-group (SDRG) method~\citep{SDRG1}
(for a review, see Refs.~\onlinecite{SDRG2,igloi-monthus-review2-arxiv}).

Given the importance of the infinite-randomness concept, it is desirable
to study it through other unbiased methods. The Monte Carlo method
was shown to be up to the task.~\citep{pich-etal-prl98,shu-etal-prb16}
Evidently, it is also desirable to use the DMRG method since it is
suitable for ground-state quantities and can be used to study systems
plagued by the sign problem. The earlier attempts were either controversial~\citep{hamacher}
or restricted to small systems~\citep{hidaale} (see also Ref.~\onlinecite{ruggiero-etal-prb16}).
More recently, tensor network based methods were developed.~\citep{rommerTNT,evenblyale}

In this work, we present an alternative DMRG algorithm (we call it
adaptive DMRG) for disordered systems which is capable of improving
the stability of the DMRG for relatively high degrees of disorder
and able to reach comparatively large systems when compared to the
conventional algorithm. We will apply our method to the random spin-1/2
chain in order to benchmark our algorithm and subsequently to the
random bilinear-biquadratic spin-1 chain where we find new results
for the correlation function, which is also confirmed by strong-disorder
renormalization-group calculations.

This work is organized as follows. In Sec.~\ref{sec:Models}, we
introduce the studied models and review some known results. In Sec.~\ref{sec:DMRG-results}
we introduce our adaptive DMRG method comparing it with either exact
diagonalization (when possible) or the standard DMRG method. In Sec.~\ref{sec:SDRG-RESULTS},
we present our SDRG calculations confirming the new DMRG results on
the spin-1 chain and generalizing it to other systems. Finally, we
report our conclusions in Sec.~\ref{sec:CONCLUSIONS}.

\section{\label{sec:Models}Models and some known results}

\subsection{The random antiferromagnetic spin-1/2 XXZ chain}

The random antiferromagnetic spin-1/2 XXZ chain is described by the
Hamiltonian

\begin{equation}
H=\sum_{i=1}^{L-1}J_{i}\left(s_{i}^{x}s_{i+1}^{x}+s_{i}^{y}s_{i+1}^{y}+\Delta s_{i}^{z}s_{i+1}^{z}\right),\label{eq:H-XXZ}
\end{equation}
where $\mathbf{s}_{i}$ are spin-1/2 operators, $\Delta$ is the system
anisotropy, and $0<J_{i}<\Omega$ are uncorrelated random couplings
distributed according to the distribution 
\begin{equation}
P(J)=\frac{D}{\Omega}\left(\frac{\Omega}{J}\right)^{1-1/D}.\label{eq:P(J)}
\end{equation}
Here, $\Omega$ sets the energy scale, and $D$ parameterizes the
disorder strength.

The model (\ref{eq:H-XXZ}) is one of the most studied random systems
exhibiting low-energy infinite-randomness physics. For $-1/2<\Delta\leq1$,
the clean Luttinger liquid is perturbatively unstable against any
amount of disorder ($D>0$) with a random-single (RS) state replacing
it as the true ground state~\citep{doty-fisher,fisher94-xxz}. The
RS state is approximately a collection of nearly independent singlet
pairs in which their size $\ell$ and excitation energy $\omega$
are related via an exotic activated scaling 
\begin{equation}
\ln\omega\sim-\ell^{\psi},\label{eq:activated}
\end{equation}
with universal tunneling exponent $\psi=\frac{1}{2}$. A striking
hallmark of the infinite-randomness character of the RS ground-state
is that the typical and arithmetic mean spin-spin correlation functions
are completely different from each other in the long-distance $\ell\gg1$
regime: while the former decays as a stretched exponential, i.e.,
$\ln C_{\text{typ}}^{\alpha}(\ell)=\overline{\ln\left|\left\langle s_{i}^{\alpha}s_{i+\ell}^{\alpha}\right\rangle \right|}\sim-\ell^{\psi_{\alpha}},$
with $\psi_{\alpha}=\psi=\frac{1}{2}$ for $\alpha=x,\ y,\ z$, the
latter decays only algebraically 
\begin{equation}
C_{\text{av}}^{\alpha}(\ell)=\overline{\left\langle s_{i}^{\alpha}s_{i+\ell}^{\alpha}\right\rangle }=\left(-1\right)^{\ell}\ell^{-\phi_{\alpha}},\label{eq:Cav}
\end{equation}
with $\phi_{\alpha}=\phi=2$. Here, $\left\langle \cdots\right\rangle $
and $\overline{\cdots}$ denote the ground-state and disorder averages,
respectively. The RS state also exhibits an emergent SO(2)$\rightarrow$SU(2)
symmetry characterized by the symmetric exponents $\psi_{\alpha}$
and $\phi_{\alpha}$: a general feature of strongly disordered SO($N$)-symmetric
antiferromagnetic spin chains.~\citep{quito-etal-07,quito-etal-07b}

It is well known that (\ref{eq:H-XXZ}) can be mapped to a chain of
interacting spinless fermions.~\citep{lieb-schultz-mattis} For the
special case $\Delta=0$, the fermions are noninteracting and thus,
large systems can be studied via exact diagonalization. For this reason,
we will use the disordered XX chain to provide benchmark results.

Another important quantity in our investigation is the entanglement
entropy (EE) which is given by 
\begin{equation}
{\cal S}(\ell)=-\text{Tr}\rho_{A}\ln\rho_{A},
\end{equation}
where $\rho_{A}$ is the zero-temperature reduced density matrix of
a continuous subsystem $A$ of size $\ell$ obtained by tracing out
the degrees of freedom of the complementary and continuous subsystem
$B$ (of size $L-\ell$). For $1\ll\ell\ll L$, it was shown in the
clean case that,~\citep{cardyentan,cvidal,prlkorepin}

\begin{equation}
{\cal S}(\ell)=\frac{c}{3\eta}\ln\ell+a^{\eta},\label{eq:EE-clean}
\end{equation}
where $c=1$ is the central charge, a is a non-universal constant,
and $\eta=1$($2$) for the systems with periodic (open) boundary
conditions. While the EE of clean chains are quite well understood,~\citep{revfazio,RMP82-277,entroreviewcalabrese}
much less is known for the case of disordered systems, which are not
conformally invariant. In particular, for the disordered antiferromagnetic
spin-1/2 Heisenberg chains it was shown~\citep{Mooreale,laflorencie-entanglement,abel-ale,calabreserandom}
that the average EE behaves very similarly to the clean system with
$\overline{{\cal S}}\sim\frac{c_{\text{eff}}}{3\eta}\ln\ell$, where
the effective central charge is given by $c_{\text{eff}}=\ln2$.

\subsection{The random antiferromagnetic spin-1 chains}

The other model we are interested in is the disordered spin-1 bilinear-quadratic
chain the Hamiltonian of which is

\begin{equation}
H=\sum_{i=1}^{L-1}J_{i}\left[\cos\theta\mathbf{S}_{i}\cdot\mathbf{S}_{i+1}+\sin\theta\left(\mathbf{S}_{i}\cdot\mathbf{S}_{i+1}\right)^{2}\right],\label{eq:H-S1}
\end{equation}
where $\mathbf{S}_{i}$ are spin-1 operators, $J_{i}>0$ are random
independent couplings distributed according to Eq.~(\ref{eq:P(J)}),
and $\theta$ is an angle parametrizing the ``anisotropy'' between
the bilinear and the biquadratic terms. The zero-temperature phase
diagram of this model was shown to be very rich,~\citep{quito-hoyos-miranda-prl15}
exhibiting six phases: a ferromagnetic phase, a Mesonic RS phase,
a Baryonic RS phase, a Haldane phase, a Griffiths phase, and a Large
Spin phase. Interestingly, and like the XXZ spin-1/2 chain, all the
RS phases were shown to have an emergent SU(3) symmetry out of an
SO(3) symmetric chain.~\citep{quito-etal-07,quito-etal-07b} As in
the random spin-1/2 antiferromagnetic chain, the emergent SU(3) symmetry
is manifest in all correlation functions. Let $\Lambda^{\alpha}$
($\alpha=1,\,\dots,\,8$) be the eight generators of the fundamental
representation of the SU(3) group, which can be chosen as: $\Lambda^{1}=S^{x}$,
$\Lambda^{2}=S^{y}$, $\Lambda^{3}=S^{z}$, $\Lambda^{4}=S^{x}S^{y}+S^{y}S^{x}$,
$\Lambda^{5}=S^{x}S^{z}+S^{z}S^{x}$, $\Lambda^{6}=S^{y}S^{z}+S^{y}S^{x}$,
$\Lambda^{7}=\left(S^{x}\right)^{2}-\left(S^{y}\right)^{2}$, and
$\Lambda^{8}=\frac{1}{\sqrt{3}}\left[3\left(S^{z}\right)^{2}-2\right]$.
Therefore, the arithmetic average correlation function 
\begin{equation}
C_{\text{av}}^{\alpha}(x)=\overline{\left\langle \Lambda_{i}^{\alpha}\Lambda_{i+x}^{\alpha}\right\rangle },\label{eq:c-alfa}
\end{equation}
decays with the universal and isotropic exponent $\phi_{\alpha}=\phi$.
Likewise, the typical correlation functions also decay as a stretched
exponential with exponent $\psi_{\alpha}=\psi$.

The Mesonic SU(3) RS phase was shown to have similar correlations
as the SU(2) RS phase. Actually, all Mesonic SU($N$) RS phases share
the same long-distance behavior with the exponents of the typical
and the mean correlation being $\phi_{\alpha}=\psi_{\alpha}^{-1}=2$.~\citep{netoSUNdesor}
On the other hand, it was shown in Ref.~\onlinecite{netoSUNdesor}
that $\psi_{\alpha}^{-1}=N$ for Baryonic SU($N$) RS phases. In addition,
based on some assumptions, it was argued that $\phi_{\alpha}=4/N$.
However, as shown later in Sec.~\ref{sec:SDRG-RESULTS} and confirmed
by our DMRG results in Sec.~\ref{sec:DMRG-results}, one of the assumptions
does not hold and, as a novel result of this work, the correct result
is $\phi_{\alpha}=2$ independent of the symmetry group rank.

\section{DMRG study\label{sec:DMRG-results}}

In this section, we show how the standard application of the DMRG
technique fails in describing the strongly disordered quantum systems
(\ref{eq:H-XXZ}), and then introduce our adaptive DMRG strategy in
order to remedy this situation.

\subsection{The antiferromagnetic XX and Heisenberg spin-1/2 chains}

Let us start with the random XXZ antiferromagnetic spin-1/2 chain
(\ref{eq:H-XXZ}). We first focus on the free fermionic case $\Delta=0$
and then on the SU(2) symmetric case $\Delta=1$.

\begin{figure}[!t]
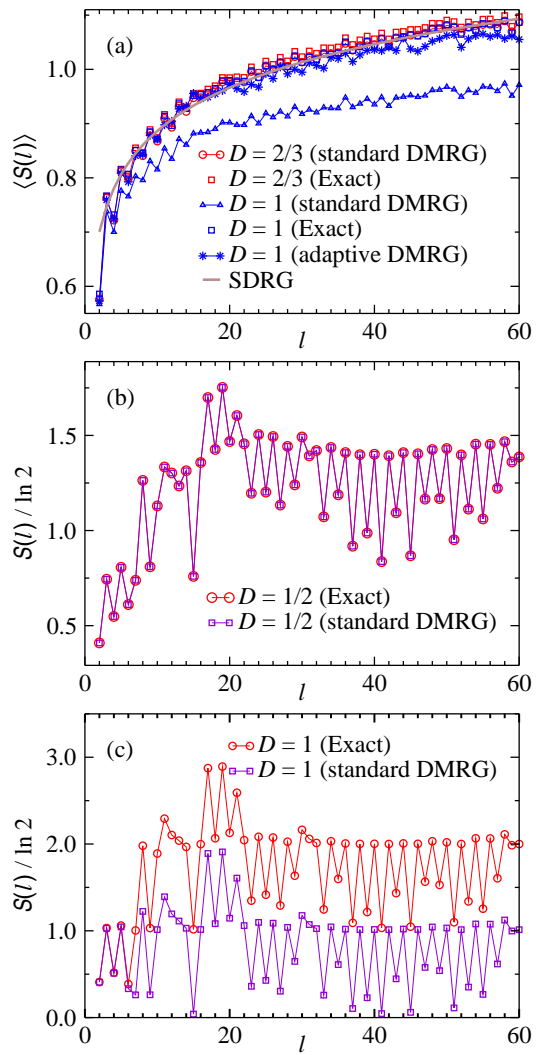

\begin{centering}
\includegraphics[clip,width=0.8\columnwidth]{fig1a}\\
 \smallskip{}
 \includegraphics[clip,width=0.8\columnwidth]{fig1b}\\
 \smallskip{}
 \includegraphics[clip,width=0.8\columnwidth]{fig1c} 
\par\end{centering}

\caption{\label{fig:1} The EE $S$ as a function of the subsystem size $l$
for the random XX chain obtained via exact diagonalization and via
the standard and adaptive DMRG methods. We have considered chains
of size $L=120$ and different values of disorder strength $D$ (see
legends). In panel (a) the EE is averaged over $2\,000$ disorder
realizations. In panels (b) and (c), the EE is computed for a single
disorder realization. The thick brown line in panel (a) corresponds
to the curve $\frac{\ln2}{6}\ln l+0.62$.}
\end{figure}

First, we report being able to obtain the ground-state energy of the
disordered XX chain with high accuracy by using the standard DMRG.
For chains of sizes $L=120$ and considering $m\sim200$ states in
the DMRG truncation,~\citep{white} we found that the errors in the
energies are typically smaller than $\sim10^{-10}$ and the discarded
weights are $\lesssim10^{-10}$. Having accomplished this, we would
expect to obtain accurate results for the EE, as well. Comparing with
the exact EE obtained via the free-fermion map,~\citep{peschel-jpa03,laflorencie-entanglement}
this is indeed the case for system sizes $L=120$ and disorder $D\lesssim\frac{2}{3}$
as shown in Figs.~\hyperref[fig:1]{\ref{fig:1}(a)} and \hyperref[fig:1]{(b)}
where, respectively, we study the average EE and the EE of a single
chain. On the other hand, for stronger disorder $D\gtrsim1$, surprisingly,
we verified that the standard DMRG algorithm fails to correctly describe
the EE as explicit in Figs.~\hyperref[fig:1]{\ref{fig:1}(a)} and
\hyperref[fig:1]{(c)}. We also note that the average EE changes very
little when the number of states increases from $m=160$ to $m=260$.
For further comparison, we also plot the average EE $\left\langle {\cal S}(\ell)\right\rangle \sim\frac{1}{6}c_{\text{eff}}\ln(\ell)$
with universal $c_{\text{eff}}=\ln2$ as predicted by the strong-disorder
RG method.~\citep{Mooreale}

\subsubsection*{The adaptive DMRG method}

We now provide the basic notion behind our adaptive DMRG method. In
Fig.~\hyperref[fig:1]{\ref{fig:1}(c)} we present the EE for a specific
coupling configuration \{$J_{1},\,J_{2},\,...,\,J_{L-1}\}$ distributed
according to Eq.~(\ref{eq:P(J)}) with $D=1$. Clearly, the standard
DMRG fails to reproduce the exact result for $\ell>6$. It turns out
that, for this specific disorder realization, spins $7$ and $78$
are strongly entangled and locked into a singlet state to a very high
degree of approximation, as predicted by the SDRG method (see Refs.~\onlinecite{hoyos-rigolin,getelina-etal-2017}
for a precise quantification of this statement). As a consequence
of the activated dynamics (\ref{eq:activated}), its effective excitation
energy can be smaller than the standard DMRG error, which we have
set as $\sim10^{-10}$. In that case, the standard DMRG method could
easily get stuck in an excited/metastable state and miss the $\approx\ln2$
contribution of that singlet pair to the EE for $\ell>6$. \footnote{A similar situation may also happen in frustrated systems, where there
are several states with energies very close to the ground energy.}

Is it possible to recover the missing pair? As we mentioned before,
increasing the number of states does not help. Here, we suggest an
alternative route which works in most cases. Lowering the disorder
while maintaining roughly the same realization (as explained below),
the excitation gap between spins $7$ and $78$ increases, and thus,
the standard DMRG method should correctly describe the EE. This is
exactly the case as verified in Fig.~\hyperref[fig:1]{\ref{fig:1}(b)}.
There, we considered the same coupling configuration as in Fig.~\hyperref[fig:1]{\ref{fig:1}(c)}
but with the square root taken: $\{\sqrt{J_{1}},\,\sqrt{J_{2}},\,...,\,\sqrt{J_{L-1}}\}$,
which is equivalent to having the coupling constants distributed according
to Eq.~(\ref{eq:P(J)}) with $D=\frac{1}{2}$. A caveat is in order
here. Notice that, for the XX spin-1/2 chain, the SDRG method predicts
the same RS state for chains in Figs.~\hyperref[fig:1]{\ref{fig:1}(b)}
and \hyperref[fig:1]{(c)}. Evidently, there are stronger corrections
to the RS state for smaller $D$.~\citep{hoyos-rigolin,getelina-etal-2017}

Given the possibility of capturing the correct ground state for weaker
disorder strength, we then propose the following adaptive DMRG strategy.
We start with a weakly disordered chain (say, with disorder strength
$D_{0}$) where the standard DMRG method is successful. After obtaining
\LyXZeroWidthSpace \LyXZeroWidthSpace the quasi-exact Eigenstate $\left|\Psi_{D_{0}}\right\rangle $,
we use it as the initial guess in the Lanczos or Davidson procedure
for the new disorder strength $D_{0}+\delta D$ (where the new couplings
are simply $J_{i}^{1+\delta D/D_{0}}$). For $\delta D\ll D_{0}$,
we expect $\left|\Psi_{D_{0}}\right\rangle $ to be a very good starting
point for $\left|\Psi_{D_{0}+\delta D}\right\rangle $. Here, we need
to use the step-to-step wave function transformation during the sweeps
as described in Ref.~\onlinecite{dmrg2dwhite}. \LyXZeroWidthSpace{}
We perform a few (about $4$) sweeps in order to obtain the new quasi-exact
Eigenstate $\left|\Psi_{D_{0}+\delta D}\right\rangle $. Finally,
we then iterate this procedure until the desired disorder strength
$D$ is reached. Since the DMRG is able to obtain the quasi-exact
states for small disorder strengths, by using the above procedure
the DMRG will adiabatically adapt a new basis to represent the new
eigenstates.~\citep{tDMRG-white} If there is no abrupt change in
the energy levels (as a function of the disorder strength), \LyXZeroWidthSpace{}
it is then expected that the above procedure will find the true (low-energy)
states and will not get stuck in metastable states. As we show in
the following, this is indeed the case.

Our strategy certainly may sound numerically costly. However, notice
that in many cases it is desirable to study many different disorder
strengths $D$. Our strategy becomes a natural one when this is the
case.

\begin{figure}[!b]
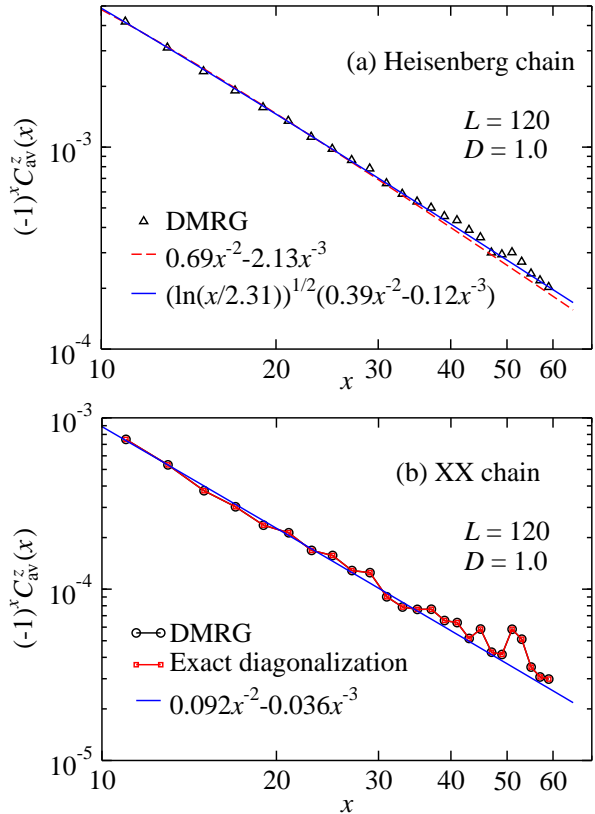

\begin{centering}
\includegraphics[clip,width=0.9\columnwidth]{fig2a}\\
\smallskip{}
\includegraphics[clip,width=0.9\columnwidth]{fig2b}
\par\end{centering}

\caption{\label{fig:2} Log-log plot of the arithmetic average correlation
function $C_{\text{av}}^{z}(x)$ vs. $x$ for (a) the Heisenberg and
(b) the XX chains. The system size $L=120$, the disorder strength
$D=1$ and we averaged over $1\,000$ disorder realizations. The DMRG
data is obtained using the adaptive strategy. The solid and dashed
lines are best fit of the DMRG data.}
\end{figure}

As a demonstration, we shown in Fig.~\hyperref[fig:1]{\ref{fig:1}(a)}
the arithmetic average EE obtained using our adaptive strategy starting
from $D_{0}=0.4$ and increasing it in steps of $\delta D=0.06$ until
we reach $D=1$. We observed that for this sequence of $D$'s the
adaptive DMRG algorithm is able to reproduce the exact EE for almost
all chains for $D=1$ and $L=120$. As expected, we verified that
decreasing the value of $\delta D$ improves the adaptive DMRG method
with the associated increase in CPU time.

Let us now discuss the spin-spin correlations (\ref{eq:Cav}). In
order to avoid border effects, we measure $\left\langle S_{i}^{z}S_{j}^{z}\right\rangle $
in the center part of the chain by considering only $\frac{1}{4}L<i<j<\frac{3}{4}L$.
The disorder average is performed over all possible distances $x=j-i$
within that range and over various different disorder realizations.

In Fig.~\ref{fig:2} we present the adaptive DMRG results for $C_{\text{av}}(x)$
for the random spin-1/2 XX and the Heisenberg chain for systems of
size $L=120$, disorder strength $D=1,$ and $10^{3}$ disorder realizations.
Our results are in perfect agreement with analytical and previous
numerical results in which the decay exponent is $\phi=2$ for both
models.~\citep{fisher94-xxz,girvenale,laflorencie-ale} For the Heisenberg
model, it is interesting to contrast this with the clean exponent
$\phi_{\text{clean}}=1$.~\citep{lutherpeschel2} Recently, a Quantum
Monte Carlo study proposed a logarithmic correction to the correlation
function for the Heisenberg model.~\cite{shu-etal-prb16} It is not
within the scope of the present work to further investigate this feature
which would require longer chains and better statistics. Here, we
simply report that our data are also compatible with it as shown in
Fig.~\hyperref[fig:1]{\ref{fig:1}(a)}.

\subsection{The disordered bilinear-biquadratic spin-1 chain}

We now present our DMRG study on the random spin-1 chain Eq.~(\ref{eq:H-S1}).
Our purpose is to use our adaptive DMRG strategy in a strongly disordered
system which is not in the well-studied SU(2) infinite-randomness
universality class. We then focus on the case $\theta=\frac{\pi}{4}$
which exhibits exact SU(3) symmetry {[}i.e., the Hamiltonian (\ref{eq:H-S1})
becomes $H=\sum_{i}J_{i}\boldsymbol{\Lambda}_{i}\cdot\boldsymbol{\Lambda}_{i+1}+\text{const}${]}
placing the system in the Baryonic RS phase.~\citep{quito-hoyos-miranda-prl15}

\begin{figure}[!b]
\begin{centering}
\includegraphics[clip,width=0.9\columnwidth]{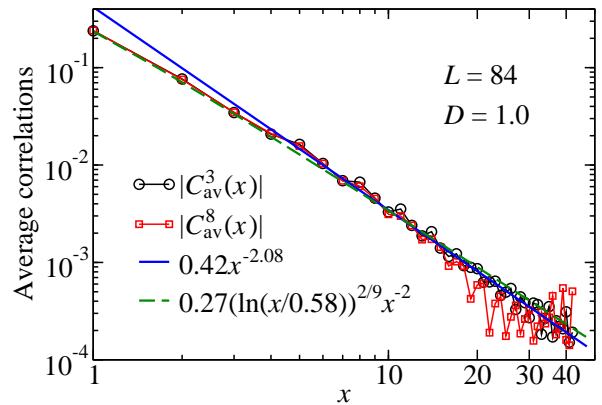} 
\par\end{centering}

\caption{\label{fig:3} The arithmetic average correlation functions $C_{\text{av}}^{3}$
and $C_{\text{av}}^{8}$ {[}see Eq.~(\ref{eq:c-alfa}){]} for the
SU(3)-symmetric disordered spin-1 bilinear-biquadratic chain Eq.~(\ref{eq:H-S1})
with $\theta=\frac{\pi}{4}$, system size $L=84$, and disorder strength
$D=1$. The continuous blue and dashed green lines are the best fit
of $C_{\text{av}}^{3}$ for $x>10$. They are compatible with the
SDRG prediction $C_{\text{av}}\sim x^{-2}$. As in the spin-$1/2$
case, a logarithmic is also compatible with our data. The DMRG data
are obtained using the adaptive strategy and averaging over $1\,000$
disorder realizations.}
\end{figure}

In Fig.~\ref{fig:3}, we plot the arithmetic average correlations
$C_{\text{av}}^{\alpha}(x)$ Eq.~(\ref{eq:c-alfa}) for $\alpha=3$
and $8$, $D=1$ and $L=84$ . The average was performed similarly
to the spin-1/2 case considering all the spin pairs $S_{i}$ and $S_{j}$
in the range $\frac{1}{4}L<i<j<\frac{3}{4}L$. We verify that $C_{{\rm av}}^{\alpha}\sim x^{-\phi}$
with $\phi=2$ is consistent with our numerical data. This is a novel
result which is in agreement with the predictions of the SDRG method
of Sec.~\ref{sec:SDRG-RESULTS}. It is interesting to compare with
the clean chain exponent $\phi_{\text{clean}}=\frac{4}{3}$.~\citep{itoi-biqua-clean}
Similarly to the spin-1/2 case, the logarithmic correction of the
clean system {[}$\propto\ln^{2/9}(x)${]}~\citep{itoi-biqua-clean}
is also compatible with our data. We report that similar results were
also obtained considering other system sizes $32\leq L\leq84$ and
disorder strengths $\frac{1}{4}\leq D\leq1$. In addition, we report
that (not shown) $C_{{\rm av}}^{\alpha}(x)$ oscillates with period
of $3$, as a consequence of the antiferromagnetic SU(3)-symmetric
character of the ground state.~\citep{quito-hoyos-miranda-prl15}
As expected, we observed that both correlations are identical within
the DMRG error.

\section{Strong-disorder RG study\label{sec:SDRG-RESULTS}}

In this section we compute the arithmetic average correlation function
$C_{\text{av}}^{\alpha}$ {[}see Eq.~(\ref{eq:c-alfa}){]} for the
spin-1 chain (\ref{eq:H-S1}) in the strong-disorder limit and in
the phase of emergent Hadronic SU(3) symmetry. For that reason, we
will employ the strong-disorder renormalization-group (SDRG) method
developed in Refs.~\onlinecite{netoSUNdesor,quito-hoyos-miranda-prl15}.

\subsection{The SU(3) random-singlet ground state\label{sub:The-SU(3)-random-singlet}}

\begin{figure}[t]
\begin{centering}
\includegraphics[clip,width=0.9\columnwidth]{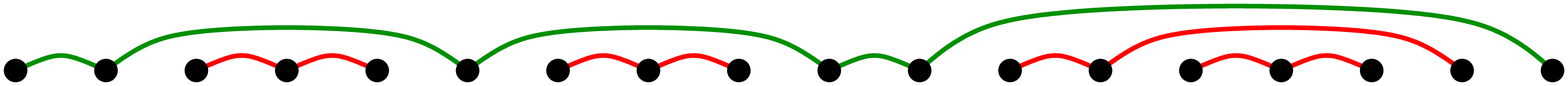} 
\par\end{centering}

\caption{Sketch of the SU(3) random-singlet state. Sites connected by links
are in a singlet state formed by $3,\,6,\,9,\,\dots$ spins. Notice
the links do not overlap. Different colors represent singlets with
different number of spins. \label{fig:GS}}
\end{figure}

For strong disorder strength $D\gg1$ (and very plausibly for any
$D>0$), the ground state of the Hamiltonian (\ref{eq:H-S1}) is the
SU(3) random singlet state for $\frac{\pi}{4}\leq\theta<\frac{\pi}{2}$.~\citep{quito-hoyos-miranda-prl15}
In this case, due to the emergent SU(3) symmetry, the ground-state
is composed by nearly independent SU(3) singlets as sketched in Fig.~\ref{fig:GS}.

Unlike the usual SU(2) spin-1/2 random-singlet state where all singlets
are made of spin pairs, in the SU(3) case they can be made of any
multiple of three spins. Interestingly, it has been shown that the
clustering of spins disentangles from the chain energetics near the
infinite-randomness fixed point.~\citep{netoSUNdesor} Therefore,
the ground state depicted in Fig.~\ref{fig:GS} can be obtained in
the following simple fashion: (i) one randomly chooses a neighboring
spin pair in the chain and (ii) fuses them together in a new effective
spin (a new spin cluster). (ii.a) If the total number of original
spins in the new cluster is a multiple of three, the cluster is removed
from the system since they form a singlet as in Fig.~\ref{fig:GS},
otherwise, (ii.b) it remains in the system ``waiting'' for a new
decimation. The procedure (i) and (ii) is iterated until all spins
become clustered into singlets (assuming that the lattice size is
a multiple of three) as in Fig.~\ref{fig:GS}.

With these simplified clustering rules, it is possible to compute
the probability that two original spins $\ell$ lattice sites apart
become clustered in the same singlet. Assuming that they share correlations
of order unity, then $C_{\text{av}}^{\alpha}$ would simply be proportional
to that probability, since spins in different singlets would have
exponentially small correlation. In this way, it was concluded in
Ref.~\onlinecite{netoSUNdesor} that $C_{\text{av}}^{\alpha}\sim\ell^{-\frac{4}{N}}$,
with $N=3$. (This generalizes to all SU($N$) random singlet states
where singlets are composed by multiples of $N$ original spins).

However, as we show below, the assumption that spins belonging to
the same singlet have strong correlations is not correct. Therefore,
we need a better understanding of the many possible singlet states
in order to correctly compute $C_{\text{av}}^{\alpha}$.

\subsection{Correlations in the SU(3) singlets}

\begin{figure}[b]
\begin{centering}
\includegraphics[clip,width=0.45\columnwidth]{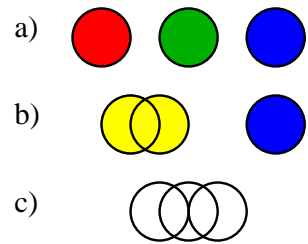} 
\par\end{centering}

\caption{Schematic representation of the clustering process of three spins
into a singlet state. Colors are for aesthetic purposes only.\label{fig:s3}}
\end{figure}

The simplest and most common SU(3) singlet is the one made of three
spins (see Fig.~\ref{fig:s3}). It can be readily obtained by the
anti-symmetrization of the three possible spin flavors {[}corresponding
to Fig.~\hyperref[fig:s3]{\ref{fig:s3}(c)}{]}: 
\begin{align}
\left|s_{\text{3-spins}}\right\rangle = & \frac{1}{\sqrt{6}}\left(\left|1,0,-1\right\rangle +\left|0,-1,1\right\rangle +\left|-1,1,0\right\rangle \right.\nonumber \\
 & \left.-\left|0,1,-1\right\rangle -\left|1,-1,0\right\rangle -\left|-1,0,1\right\rangle \right).\label{eq:singlet-3}
\end{align}
It is then clear that any spin pair $i,j$ in the $\left|s_{\text{3-spins}}\right\rangle $
singlet state share correlation of order unity, namely, $C_{i,j}^{\alpha}=\left\langle s_{\text{3-spins}}\left|\Lambda_{i}^{\alpha}\Lambda_{j}^{\alpha}\right|s_{\text{3-spins}}\right\rangle =-\frac{1}{3}$,
for any $\alpha$.

Another way of obtaining $\left|s_{\text{3-spins}}\right\rangle $
is by following the SDRG method.~\citep{netoSUNdesor,quito-hoyos-miranda-prl15}
One first fuses, say, spins $S_{1}$ and $S_{2}$ into a new spin-1
effective degree of freedom $\tilde{S}$ {[}corresponding to Fig.~\hyperref[fig:s3]{\ref{fig:s3}(b)}{]},
which is then decimated with spin $S_{3}$ into a singlet. With respect
to the original flavors, the $\tilde{S}$ degrees of freedom are 
\begin{align}
\left|\tilde{1}\right\rangle  & =\frac{1}{\sqrt{2}}\left(\left|1,0\right\rangle -\left|0,1\right\rangle \right),\nonumber \\
\left|\tilde{0}\right\rangle  & =\frac{1}{\sqrt{2}}\left(\left|1,-1\right\rangle -\left|-1,1\right\rangle \right),\label{eq:S1-effective}\\
\left|-\tilde{1}\right\rangle  & =\frac{1}{\sqrt{2}}\left(\left|0,-1\right\rangle -\left|-1,0\right\rangle \right),\nonumber 
\end{align}
which are obtained by projecting $\mathbf{S}_{1}+\mathbf{S}_{2}$
on the triplet manifold. The state (\ref{eq:singlet-3}) is then obtained
by projecting $\tilde{\mathbf{S}}+\mathbf{S}_{3}$ on the singlet
manifold $\bar{S}=0$, i.e., 
\begin{equation}
\left|s_{\text{3-spins}}\right\rangle =\frac{1}{\sqrt{3}}\left(\left|\tilde{1},-1\right\rangle -\left|\tilde{0},0\right\rangle +\left|-\tilde{1},1\right\rangle \right).\label{eq:singlet-3b}
\end{equation}
We now ask, for instance, how $C_{1,3}^{z}$ can be obtained given
the knowledge of the singlet state (\ref{eq:singlet-3b}). First,
we notice that the correlation 
\begin{equation}
\left\langle \tilde{S}^{z}S_{3}^{z}\right\rangle =\frac{1}{6}\left\langle \bar{S}^{2}-\tilde{S}^{2}-S_{3}^{2}\right\rangle =-\frac{2}{3}.\label{eq:SS-singlet}
\end{equation}
Then, we make use of the Wigner-Eckart theorem. Since $\tilde{\mathbf{S}}$
is simply $\mathbf{S}_{1}+\mathbf{S}_{2}$ projected on the triplet
manifold, then $\mathbf{S}_{1}=c_{\tilde{S},S_{1},S_{2}}\tilde{\mathbf{S}}$.
Since we will need to deal only with the case $S_{1}=S_{2}=\tilde{S}=1$,
we lighten the notation by $c_{\tilde{S},S_{1},S_{2}}=c=\frac{1}{2}$
which can be obtained by projecting $\mathbf{S}_{1,2}$ in the multiplet
(\ref{eq:S1-effective}). Finally, we have that 
\begin{equation}
\left\langle S_{1,2}^{\alpha}S_{3}^{\beta}\right\rangle =c\left\langle \tilde{S}^{\alpha}S_{3}^{\beta}\right\rangle =-\frac{\delta_{\alpha,\beta}}{3}.
\end{equation}
We now ask about the correlations between $S_{1}$ and $S_{2}$. For
instance, 
\begin{equation}
\left\langle S_{1}^{z}S_{2}^{z}\right\rangle =\frac{1}{6}\left\langle \tilde{S}^{2}-S_{1}^{2}-S_{2}^{2}\right\rangle =-\frac{1}{3}.\label{eq:SS-triplet}
\end{equation}
With these results, we recover that $C_{i,j}^{\alpha}=\left\langle s_{\text{3-spins}}\left|\Lambda_{i}^{\alpha}\Lambda_{j}^{\alpha}\right|s_{\text{3-spins}}\right\rangle =-\frac{1}{3}$,
since it can be verified that the three-spin singlet is also an SU(3)
singlet.

\begin{figure}[t]
\begin{centering}
\includegraphics[clip,width=0.8\columnwidth]{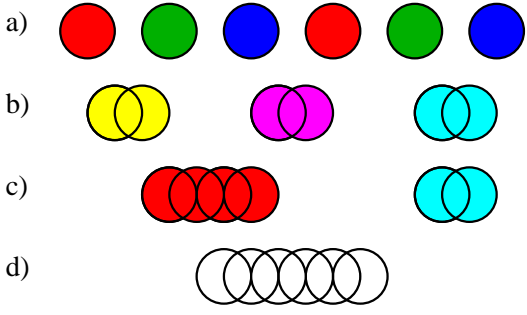} 
\par\end{centering}

\caption{Schematic representation of the clustering process of six spins into
a singlet state. Colors are for aesthetic purposes only.\label{fig:s6}}
\end{figure}

Before generalizing these results to other singlets, let us examine
the case of singlets composed by 6 spins. They can be formed in many
different ways. For our purpose, let us examine only the case in which
spins $S_{1}$ and $S_{2}$ are fused together in the new effective
spin $\tilde{S}_{1}$. Likewise spins $S_{3}$ and $S_{4}$ ($S_{5}$
and $S_{6}$) become locked into the new effective spin $\tilde{S}_{2}$
($\tilde{S}_{3}$) (see Fig.~\ref{fig:s6}). The resulting singlet
is obtained by anti-symmetrizing the effective flavors of $\tilde{S}_{1}$,
$\tilde{S}_{2}$ and $\tilde{S}_{3}$, just as in the three-spin case,
resulting in the singlet $\left|s_{\text{6-spins(a)}}\right\rangle $
given by (\ref{eq:singlet-3}) with the flavors $m$ replaced by $\tilde{m}$
in (\ref{eq:S1-effective}) {[}corresponding to Fig.~\hyperref[fig:s6]{\ref{fig:s6}(d)}{]}.
Less straightforwardly, we can fuse spins $\tilde{S}_{1}$ and $\tilde{S}_{2}$
into the new effective spin-1 $\tilde{\tilde{S}}$ {[}the flavors
of which are given by (\ref{eq:S1-effective}) with $\tilde{m}\rightarrow\tilde{\tilde{m}}$
and $m\rightarrow\tilde{m}${]}, and then fuse $\tilde{\tilde{S}}$
with $\tilde{S}_{3}$ into the $\bar{S}=0$ singlet state $\left|s_{\text{6-spins(a)}}\right\rangle $
{[}corresponding to Figs.~\hyperref[fig:s3]{\ref{fig:s3}(c)} and
\hyperref[fig:s3]{(d)}{]} given by (\ref{eq:singlet-3b}) with with
$\tilde{m}\rightarrow\tilde{\tilde{m}}$ and $m\rightarrow\tilde{m}$
(as before).

Let us now compute the correlations. Consider for instance $C_{1,2}^{z}=\left\langle S_{1}^{z}S_{2}^{z}\right\rangle =\left\langle s_{\text{6-spins(a)}}\left|S_{1}^{z}S_{2}^{z}\right|s_{\text{6-spins(a)}}\right\rangle $.
Although the singlet state is a different one, the correlation is
just as in the three-spin case (\ref{eq:SS-triplet}) yielding $C_{1,2}^{\alpha}=C_{3,4}^{\alpha}=C_{5,6}^{\alpha}=-\frac{1}{3}$
for any $\alpha$ (due to symmetry). Hence, as in the three-spin singlet
case, there are strong correlations. Notice this strong correlation
is a general feature when two original spins are decimated together
into an $\tilde{S}=1$ cluster. Afterwards, renormalizations involving
$\tilde{S}$ do not change the correlation between the original spins.

However, the correlations between other spin pairs are much weaker.
Consider for instance $C_{1,3}^{z}$. Making use of the Wigner-Eckart
theorem, then $\left\langle S_{1}^{z}S_{3}^{z}\right\rangle =c^{2}\left\langle \tilde{S}_{1}^{z}\tilde{S}_{2}^{z}\right\rangle =-\frac{1}{3}c^{2}$,
since $\tilde{\mathbf{S}}_{1}$ and $\tilde{\mathbf{S}}_{2}$ are
fused into a $\tilde{\tilde{S}}=1$ cluster from which follows (\ref{eq:SS-triplet}).
Finally, let us compute $C_{1,5}^{z}$. We will need to compute $\left\langle \tilde{\tilde{S}}^{z}\tilde{S}_{3}^{z}\right\rangle =-\frac{2}{3}$
since they fuse into a singlet, and thus follows (\ref{eq:SS-singlet}).
From the Wigner-Eckart theorem, $\left\langle S_{1}^{z}S_{5}^{z}\right\rangle =c^{2}\left\langle \tilde{S}_{1}^{z}\tilde{S}_{3}^{z}\right\rangle =c^{3}\left\langle \tilde{\tilde{S}}^{z}\tilde{S}_{3}^{z}\right\rangle =-2c^{3}/3$.

We then conclude that, by symmetry, $C_{i,j}^{\alpha}=-\frac{1}{12}$
for all other pairs $(i,j)$ that are not ($1,2$), ($3,4$) or ($5,6$).
The important feature, as we show below, is that some longer-ranged
correlations pick up powers of $c$, and thus, can be exponentially
smaller in larger clusters.

We are now in a position to compute the correlations between spins
$S_{i}$ and $S_{j}$ belonging to a generic SU(3) singlet. Since
they belong to the same singlet cluster, they will be fused together
at some point of the SDRG flow. Let $\tilde{S}$ be the effective
cluster they first become fused together. Also, let $\tilde{S}_{i}$
and $\tilde{S}_{j}$ be the effective clusters that originated $\tilde{S}$.
Necessarily, spin $S_{i}$ ($S_{j}$) belongs to cluster $\tilde{S}_{i}$
($\tilde{S}_{j}$). Then $\left\langle \tilde{S}_{i}^{z}\tilde{S}_{j}^{z}\right\rangle =\frac{1}{6}\left\langle \tilde{S}^{2}-\tilde{S}_{i}^{2}-\tilde{S}_{j}^{2}\right\rangle =\frac{1}{6}\left(\tilde{S}\left(\tilde{S}+1\right)-4\right)$
and $C_{i,j}^{z}=c^{k_{i}+k_{j}}\left\langle \tilde{S}_{i}^{z}\tilde{S}_{j}^{z}\right\rangle $,
where $k_{i}$ ($k_{j}$) is the number of fusions undergone by $S_{i}$
($S_{j}$) before $\tilde{S}_{i}$ is clustered with $\tilde{S}_{j}$.
Finally, by symmetry, 
\begin{equation}
C_{i,j}^{\alpha}=\frac{1}{6}\left(\tilde{S}\left(\tilde{S}+1\right)-4\right)c^{k_{i}+k_{j}},\label{eq:Cij}
\end{equation}
for any spins belonging to the same singlet cluster and $c=\frac{1}{2}$.
Recall that $\tilde{S}=0$ ($1$) when the effective clusters of $S_{i}$
and $S_{j}$ are fused together into a singlet (triplet) state.

\subsection{Mean correlation function}

Having computed the correlation between two spins belonging to the
same cluster (\ref{eq:Cij}), we now proceed to compute the arithmetic
mean correlation function (\ref{eq:c-alfa}). Following the SDRG philosophy,
spins in different singlet clusters share very weak correlations and
therefore, do not contribute to the long-distance behavior of $C_{\text{av}}^{\alpha}$
(we set $C_{i,j}^{\alpha}=0$ for $i$ and $j$ belonging to different
spin singlet clusters).

We then proceed by numerically implementing the SDRG method as explained
in the following. We focus on the SU(3)-symmetric spin chain $\theta=\frac{\pi}{4}$
in the Hamiltonian (\ref{eq:H-S1}) (but this also applies to $\frac{\pi}{4}\leq\theta<\frac{\pi}{2}$)
with coupling constants drawn from the distribution (\ref{eq:P(J)}).
We then decimate the entire chain using the SDRG rules as explained
in Refs.~\onlinecite{netoSUNdesor,quito-hoyos-miranda-prl15}.
We choose the largest coupling in the system, say, $J_{2}$, and decimate
the corresponding effective cluster spin pair either by (i) removing
them from the system (which happens when the total number of original
spins in both clusters is a multiple of three) or (ii) by clustering
them in a new effective spin-1 cluster (which happens otherwise).
In the case (i) of a singlet decimation, the neighboring spin clusters
become connected via a weaker coupling of magnitude $\tilde{J}=\frac{2J_{1}J_{3}}{9J_{2}}$.
On the other hand for the case (ii), the new couplings connecting
to the new effective spin cluster are $\tilde{J}_{1,3}=\frac{1}{2}J_{1,3}$.

\begin{figure}[!t]
\begin{centering}
\includegraphics[clip,width=0.9\columnwidth]{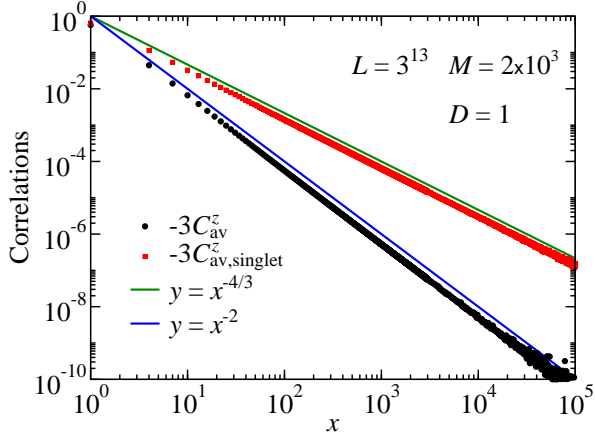} 
\par\end{centering}

\caption{The arithmetic mean spin-spin correlation $C_{\text{av}}^{\alpha}$
function as a function of the spin separation $x$ computed for distances
$x=1+3n$, with $n\in\mathbb{N}_{+}$. For comparison, we also show
the singlet spin-spin correlation $C_{\text{av,singlet}}^{\alpha}$
(see text). The system size is $L=3^{12}$ (where periodic boundary
conditions were considered), the disorder parameter is $D=1$, and
we have averaged over $M=2\times10^{3}$ different disorder realizations.
\label{fig:C-SDRG}}
\end{figure}

After the entire chain is decimated, the SDRG ground state is obtained
(see Fig.~\ref{fig:GS}) and the correlations can be computed via
(\ref{eq:Cij}). Averaging over all distances and over $M$ different
disorder realizations, $C_{\text{av}}^{\alpha}$ is obtained as shown
in Fig.~\ref{fig:C-SDRG}. For comparison, we also plot the arithmetic
mean singlet-correlation $C_{\text{av,cluster}}^{\alpha}$, which
is simply the probability of finding two spins belonging to the same
spin singlet cluster multiplied by $-\frac{1}{3}$. Notice that $C_{\text{av,cluster}}^{\alpha}(x)\sim x^{-\phi_{\text{cluster}}}$
with $\phi_{\text{cluster}}=\frac{4}{3}$ (as shown in Ref.~\onlinecite{netoSUNdesor})
and that $C_{\text{av}}^{\alpha}(x)\sim x^{-\phi}$ with $\phi=2$.
We have studied chains of different sizes and different disorder strengths
and verified the universality of these exponents.

It is desirable to obtain an analytical derivation for the universal
exponents $\phi=2$ and $\phi_{\text{cluster}}=\frac{4}{3}$. We will
learn from this quest that $C_{\text{av,{\rm cluster}}}$ is dominated
by spin pairs in large clusters composed by several original spins,
and that the exponential suppression of correlations after many clusterings
{[}see Eq.~(\ref{eq:Cij}){]} is so strong that $C_{\text{av}}$
becomes dominated by original spins pairs that become locked together
in a cluster for the first time.

We start our analysis with $Q(t;\rho)$: the probability of finding
a spin cluster composed of $t$ original spins at the length scale
$x=\rho^{-1}=L/N_{T}$ where $N_{T}$ is the total number of spin
clusters at that length scale, $\rho$ is simply the density of spin
clusters in the lattice, and $L$ is the original number of lattice
sites. As mentioned in Sec.~\ref{sub:The-SU(3)-random-singlet},
the SDRG clustering rules disentangle from the system energetics in
the later stages of the SDRG flow. In that case, the flow equation
for $Q$ becomes much simpler: 
\begin{equation}
\text{d}[L\rho Q]=\text{d}N_{\text{dec}}\left(-2Q(t;\rho)+Q\stackrel{t}{\hexstar}Q\right).\label{eq:flow-Q}
\end{equation}
The left-hand-side of Eq.~(\ref{eq:flow-Q}) is simply the change
on the number of clusters containing $t$ original spins when the
system density changes from $\rho$ to $\rho+\text{d}\rho$. $\text{d}N_{\text{dec}}$
is the corresponding total number of decimations which is related
to $\rho$ via $\text{d}[L\rho]=-\left(2p+q\right)\text{d}N_{\text{dec}}$,
with $p$ ($q=1-p$) being the probability of a (non-) singlet decimation
and $\text{d}[L\rho]$ being the corresponding change in the total
number of clusters. For the SU(3) case, $p=q=\frac{1}{2}$. Generically
for the SU($N$) case, $p=\frac{1}{N-1}$. Recall that for each singlet-like
decimation, two clusters are removed while for a non-singlet decimation,
two clusters are removed but a new one is inserted. The first term
on the right-hand-side of Eq.~(\ref{eq:flow-Q}) accounts for the
removal of the two decimated clusters in every decimation. The last
term accounts for the insertion of the new cluster containing the
total number of spins: $Q\stackrel{t}{\hexstar}Q=\sum_{t_{1},t_{2}}Q(t_{1})Q(t_{2})\delta_{t,t_{1}+t_{2}}(1-\delta_{t,N}-\delta_{t,2N}-\dots)$.
The term inside the parentheses ensures that only non-singlet decimations
contribute. In order to keep the analysis simple, from now on we will
allow $Q(t)$ to be non-zero also for $t$ a multiple of $N$ and
recast this term as $qQ\stackrel{t}{\otimes}Q=q\sum_{t_{1},t_{2}}Q(t_{1})Q(t_{2})\delta_{t,t_{1}+t_{2}}$.
Exchanging $Q\stackrel{t}{\hexstar}Q$ by $qQ\stackrel{t}{\otimes}Q$
is equivalent to replacing the precise occurrence of a non-singlet
decimation by its average occurrence. Therefore, this simplification
cannot change the large-$t$ behavior of $Q(t;\rho)$, and thus, we
expect to obtain the correct value of the universal exponents $\phi$
and $\phi_{\text{cluster}}$.

\begin{figure}[t]
\begin{centering}
\includegraphics[clip,width=0.8\columnwidth]{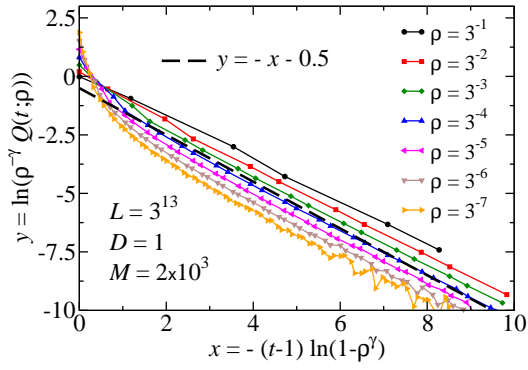} 
\par\end{centering}

\caption{The probability $Q(t;\rho)$ of finding a spin cluster composed of
$t$ original spins at the density scale $\rho$ for various different
density values. The data were obtained via the numerical implementation
of the SDRG procedure where the parameters used are the same as in
Fig.~\ref{fig:C-SDRG}. The dashed line is the simplified prediction
Eq.~(\ref{eq:Q}) with an offset for better comparison. The lines
are guide to the eyes.\label{fig:Q}}
\end{figure}

We now try a solution of type $Q(t;\rho)=A_{\rho}e^{-B_{\rho}(t-1)}$,
where $A_{\rho}$ and $B_{\rho}$ are $t$-independent functions.
From the normalization condition $\sum_{t=1}^{\infty}Q(t;\rho)=1$,
our Ansatz simplifies to $Q=A_{\rho}\left(1-A_{\rho}\right)^{t-1}$.
Plugging this result into the simplified flow equation, we find that
\begin{equation}
Q\left(t;\rho\right)=\rho^{\gamma}\left(1-\rho^{\gamma}\right)^{t-1},\label{eq:Q}
\end{equation}
where $\gamma=1-\frac{2}{N}$ and we have used the initial condition
$Q(t;1)=\delta_{t,1}$. For comparison, we plot in Fig.~\ref{fig:Q}
the probability $Q(t;\rho)$ for various different values of density
$\rho$ obtained via the numerical implementation of the SDRG procedure
as explained for Fig.~\ref{fig:C-SDRG}. As expected, the large-$t$
behavior is well described by our simplified result (\ref{eq:Q}),
although we cannot rule out a power-law correction to the exponential
dependence on $t$.

We are now able to obtain the leading behavior of $C_{\text{av,cluster}}^{\alpha}$.
This is proportional to the probability that any two original spins,
$x$ lattice sites apart, are in neighboring spin clusters at the
density scale $\rho=x^{-1}$. (We can associate $x$ with $\rho^{-1}$
because the size of the clusters is of order of the mean distance
between them.) Thus, we need the probability $R_{T}(\rho)$ that a
certain original spin is still active (i.e., belonging to some spin
cluster) at the density scale $\rho$. This is proportional to the
total number of original spins in the effective chains. Thus, $R_{T}\propto\rho\sum_{t}tQ\left(t;\rho\right)=\rho\bar{t}=\rho^{1-\gamma}$
and 
\begin{equation}
C_{\text{av,cluster}}^{\alpha}\sim\left(R_{T}(\rho)\right)^{2}\propto x^{-\phi_{\text{cluster}}},\label{eq:Cav-singlet}
\end{equation}
with $\phi_{\text{cluster}}=\frac{4}{N}$, which recovers the result
of Ref.~\onlinecite{netoSUNdesor}.

In order to compute $C_{\text{av}}^{\alpha}$, we need the probability
$R(t;\rho)=R_{T}\left(\rho\right)Q\left(t;\rho\right)$ of finding
an original spin in a cluster composed by $t$ original spins at the
density scale $\rho$. The correlations in Eq.~(\ref{eq:Cij}) are
incorporated in the following approximate way. We assume that contribution
to the correlations coming from spins $S_{i}$ and $S_{i+x}$ is $\propto c^{2k}$
where $k$ is the largest integer smaller then $\frac{t_{i}+t_{i+x}}{N}$,
where $t_{i}$ and $t_{i+x}$ are respectively the number of original
spins on the clusters containing $S_{i}$ and $S_{i+x}$ when they
are fused together at the density scale $\rho=x^{-1}$. Thus, Eq.~(\ref{eq:Cav-singlet})
is generalized to 
\begin{align}
C_{\text{av}}^{\alpha} & \sim\sum_{k=0}^{\infty}c^{2k}\sum_{kN<t_{i}+t_{j}\leq(k+1)N}R(t_{i};\rho)R(t_{j};\rho)\label{eq:Cav-SDRGa}\\
 & =\rho^{2}\left(1+\sum_{n=1}^{\infty}a_{n}\left(1-\rho^{\gamma}\right)^{n}\right)\sim ax^{-\phi},\label{eq:Cav-SDRGb}
\end{align}
where the second sum in (\ref{eq:Cav-SDRGa}) denotes a double sum
over all values $t_{i}$ and $t_{j}$ obeying the constraint $kN<t_{i}+t_{j}\leq(k+1)N$.
In the last passage, we considered only the long-distance regime $x\gg1$
where we found a universal exponent $\phi=2$ and the constant 
\begin{equation}
a=\sum_{k=0}^{\infty}\frac{c^{2k}\left((2k-1)N^{2}-N\right)}{2}=\frac{\left(N+N^{2}\right)c^{2}+N^{2}-N}{2\left(1-c^{2}\right)^{2}}.\label{eq:constant}
\end{equation}
We therefore recover the numerical SDRG results in Fig.~\ref{fig:C-SDRG}
and provide a simple theory for the DMRG results of Sec.~\ref{sec:DMRG-results}.
It is interesting to track the contributions to the constant (\ref{eq:constant}).
The exponential decay of correlations upon many projections (\ref{eq:Cij})
dictates that the main contribution comes from those spins in smaller
clusters. For this reason, the result (\ref{eq:Cav-SDRGb}) is applicable
to any SU($N$) random singlet state.

\section{Conclusions\label{sec:CONCLUSIONS}}

We have devised an adaptive density-matrix renormalization-group (DMRG)
method able to tackle strongly disordered random systems and applied
it to the random antiferromagnetic spin-1/2 chain and to the random
spin-1 with bilinear and biquadratic interactions.

The adaptive DMRG method was able to recover the known results for
the spin-1/2 chain in the literature and overcome the deficiency of
the standard DMRG method in capturing the entanglement between distant
spins in the system. For the spin-1 chain at the SU(3) symmetric point
{[}$\theta=\frac{\pi}{4}$ in (\ref{eq:H-S1}){]}, we found that the
average correlations decay as a power law with the same universal
exponent as in the spin-1/2 chains, $\phi=2$. In order to confirm
this result, we then developed a strong-disorder renormalization-group
(SDRG) framework for computing the spin-spin correlation for all SU($N$)
symmetric random-singlet states and concluded that the correlation
exponent is universal and equal to $\phi=2$ for all $N\ge2$. This
result also applies to all SO($N$)-symmetric random spin chains exhibiting
enlarged SU($N$) symmetry random-singlet ground states.~\citep{quito-etal-07,quito-etal-07b}

Our adaptive DMRG algorithm requires few changes with respect to the
standard DMRG method and thus, can be easily implemented. The input
state of our method in the high-disorder regime is self-generated
and does not rely on other methods such as those of the tensor-network-based
algorithms. Finally, the convergence of our method for larger degrees
of disorder can be controlled by setting smaller disorder increments.
Therefore, our method may be suitable to study other quantum phase
transitions driven by the disorder strength such as many-body localization
transitions. 
\begin{acknowledgments}
The authors thank D. Eloy, F. B. Ramos, and A. L. Malvezzi for providing
data for comparison, and V. L. Quito and A. Sandvik for useful discussions.
This research was supported by the Brazilian agencies FAPEMIG, FAPESP
and CNPq. J.A.H. also acknowledges the hospitality of the Aspen Center
for Physics and the financial support of NSF and Simons Foundation. 
\end{acknowledgments}

 \bibliographystyle{apsrev4-1}
\bibliography{refs_rev4}

\begin{thebibliography}{53}%
\makeatletter
\providecommand \@ifxundefined [1]{%
 \@ifx{#1\undefined}
}%
\providecommand \@ifnum [1]{%
 \ifnum #1\expandafter \@firstoftwo
 \else \expandafter \@secondoftwo
 \fi
}%
\providecommand \@ifx [1]{%
 \ifx #1\expandafter \@firstoftwo
 \else \expandafter \@secondoftwo
 \fi
}%
\providecommand \natexlab [1]{#1}%
\providecommand \enquote  [1]{``#1''}%
\providecommand \bibnamefont  [1]{#1}%
\providecommand \bibfnamefont [1]{#1}%
\providecommand \citenamefont [1]{#1}%
\providecommand \href@noop [0]{\@secondoftwo}%
\providecommand \href [0]{\begingroup \@sanitize@url \@href}%
\providecommand \@href[1]{\@@startlink{#1}\@@href}%
\providecommand \@@href[1]{\endgroup#1\@@endlink}%
\providecommand \@sanitize@url [0]{\catcode `\\12\catcode `\$12\catcode
  `\&12\catcode `\#12\catcode `\^12\catcode `\_12\catcode `\%12\relax}%
\providecommand \@@startlink[1]{}%
\providecommand \@@endlink[0]{}%
\providecommand \url  [0]{\begingroup\@sanitize@url \@url }%
\providecommand \@url [1]{\endgroup\@href {#1}{\urlprefix }}%
\providecommand \urlprefix  [0]{URL }%
\providecommand \Eprint [0]{\href }%
\providecommand \doibase [0]{http://dx.doi.org/}%
\providecommand \selectlanguage [0]{\@gobble}%
\providecommand \bibinfo  [0]{\@secondoftwo}%
\providecommand \bibfield  [0]{\@secondoftwo}%
\providecommand \translation [1]{[#1]}%
\providecommand \BibitemOpen [0]{}%
\providecommand \bibitemStop [0]{}%
\providecommand \bibitemNoStop [0]{.\EOS\space}%
\providecommand \EOS [0]{\spacefactor3000\relax}%
\providecommand \BibitemShut  [1]{\csname bibitem#1\endcsname}%
\let\auto@bib@innerbib\@empty
%</preamble>
\bibitem [{\citenamefont {Vidal}(2008)}]{Vidalmera}%
  \BibitemOpen
  \bibfield  {author} {\bibinfo {author} {\bibfnamefont {G.}~\bibnamefont
  {Vidal}},\ }\href {\doibase 10.1103/PhysRevLett.101.110501} {\bibfield
  {journal} {\bibinfo  {journal} {Phys. Rev. Lett.}\ }\textbf {\bibinfo
  {volume} {101}},\ \bibinfo {pages} {110501} (\bibinfo {year}
  {2008})}\BibitemShut {NoStop}%
\bibitem [{\citenamefont {Verstraete}\ \emph {et~al.}(2008)\citenamefont
  {Verstraete}, \citenamefont {Murg},\ and\ \citenamefont
  {Cirac}}]{verstraeteadvphys}%
  \BibitemOpen
  \bibfield  {author} {\bibinfo {author} {\bibfnamefont {F.}~\bibnamefont
  {Verstraete}}, \bibinfo {author} {\bibfnamefont {V.}~\bibnamefont {Murg}}, \
  and\ \bibinfo {author} {\bibfnamefont {J.~I.}\ \bibnamefont {Cirac}},\ }\href
  {\doibase 10.1080/14789940801912366} {\bibfield  {journal} {\bibinfo
  {journal} {Advances in Physics}\ }\textbf {\bibinfo {volume} {57}},\ \bibinfo
  {pages} {143} (\bibinfo {year} {2008})}\BibitemShut {NoStop}%
\bibitem [{\citenamefont {Newman}\ and\ \citenamefont
  {Barkema}(1999)}]{newman-MC}%
  \BibitemOpen
  \bibfield  {author} {\bibinfo {author} {\bibfnamefont {M.~E.~J.}\
  \bibnamefont {Newman}}\ and\ \bibinfo {author} {\bibfnamefont {G.~T.}\
  \bibnamefont {Barkema}},\ }\href@noop {} {\emph {\bibinfo {title} {Monte
  Carlo Methods in Statistical Physics}}}\ (\bibinfo  {publisher} {Claredon
  Press},\ \bibinfo {address} {Oxford},\ \bibinfo {year} {1999})\BibitemShut
  {NoStop}%
\bibitem [{\citenamefont {Sandvik}(2010)}]{sandvik-QMC}%
  \BibitemOpen
  \bibfield  {author} {\bibinfo {author} {\bibfnamefont {A.~W.}\ \bibnamefont
  {Sandvik}},\ }\href {\doibase 10.1063/1.3518900} {\bibfield  {journal}
  {\bibinfo  {journal} {AIP Conference Proceedings}\ }\textbf {\bibinfo
  {volume} {1297}},\ \bibinfo {pages} {135} (\bibinfo {year}
  {2010})}\BibitemShut {NoStop}%
\bibitem [{\citenamefont {White}(1992)}]{white}%
  \BibitemOpen
  \bibfield  {author} {\bibinfo {author} {\bibfnamefont {S.~R.}\ \bibnamefont
  {White}},\ }\href {\doibase 10.1103/PhysRevLett.69.2863} {\bibfield
  {journal} {\bibinfo  {journal} {Phys. Rev. Lett.}\ }\textbf {\bibinfo
  {volume} {69}},\ \bibinfo {pages} {2863} (\bibinfo {year}
  {1992})}\BibitemShut {NoStop}%
\bibitem [{\citenamefont {Schollw\"ock}(2005)}]{reviewdmrgS}%
  \BibitemOpen
  \bibfield  {author} {\bibinfo {author} {\bibfnamefont {U.}~\bibnamefont
  {Schollw\"ock}},\ }\href {\doibase 10.1103/RevModPhys.77.259} {\bibfield
  {journal} {\bibinfo  {journal} {Rev. Mod. Phys.}\ }\textbf {\bibinfo {volume}
  {77}},\ \bibinfo {pages} {259} (\bibinfo {year} {2005})}\BibitemShut
  {NoStop}%
\bibitem [{\citenamefont {Voit}(1995)}]{voit}%
  \BibitemOpen
  \bibfield  {author} {\bibinfo {author} {\bibfnamefont {J.}~\bibnamefont
  {Voit}},\ }\href {http://stacks.iop.org/0034-4885/58/i=9/a=002} {\bibfield
  {journal} {\bibinfo  {journal} {Reports on Progress in Physics}\ }\textbf
  {\bibinfo {volume} {58}},\ \bibinfo {pages} {977} (\bibinfo {year}
  {1995})}\BibitemShut {NoStop}%
\bibitem [{\citenamefont {Miranda}\ and\ \citenamefont
  {Dobrosavljevi\ifmmode~\acute{c}\else
  \'{c}\fi{}}(2005)}]{miranda-dobrosavljevic-rpp05}%
  \BibitemOpen
  \bibfield  {author} {\bibinfo {author} {\bibfnamefont {E.}~\bibnamefont
  {Miranda}}\ and\ \bibinfo {author} {\bibfnamefont {V.}~\bibnamefont
  {Dobrosavljevi\ifmmode~\acute{c}\else \'{c}\fi{}}},\ }\href
  {http://stacks.iop.org/0034-4885/68/i=10/a=R02} {\bibfield  {journal}
  {\bibinfo  {journal} {Rep. Prog. Phys.}\ }\textbf {\bibinfo {volume} {68}},\
  \bibinfo {pages} {2337} (\bibinfo {year} {2005})}\BibitemShut {NoStop}%
\bibitem [{\citenamefont {Vojta}(2006)}]{vojta-review06}%
  \BibitemOpen
  \bibfield  {author} {\bibinfo {author} {\bibfnamefont {T.}~\bibnamefont
  {Vojta}},\ }\href {\doibase 10.1088/0305-4470/39/22/R01} {\bibfield
  {journal} {\bibinfo  {journal} {J. Phys. A: Math. Gen.}\ }\textbf {\bibinfo
  {volume} {39}},\ \bibinfo {pages} {R143} (\bibinfo {year}
  {2006})}\BibitemShut {NoStop}%
\bibitem [{\citenamefont {Fisher}(1992)}]{fisher92}%
  \BibitemOpen
  \bibfield  {author} {\bibinfo {author} {\bibfnamefont {D.~S.}\ \bibnamefont
  {Fisher}},\ }\href {\doibase 10.1103/PhysRevLett.69.534} {\bibfield
  {journal} {\bibinfo  {journal} {Phys. Rev. Lett.}\ }\textbf {\bibinfo
  {volume} {69}},\ \bibinfo {pages} {534} (\bibinfo {year} {1992})}\BibitemShut
  {NoStop}%
\bibitem [{\citenamefont {Fisher}(1994)}]{fisher94-xxz}%
  \BibitemOpen
  \bibfield  {author} {\bibinfo {author} {\bibfnamefont {D.~S.}\ \bibnamefont
  {Fisher}},\ }\href {\doibase 10.1103/PhysRevB.50.3799} {\bibfield  {journal}
  {\bibinfo  {journal} {Phys. Rev. B}\ }\textbf {\bibinfo {volume} {50}},\
  \bibinfo {pages} {3799} (\bibinfo {year} {1994})}\BibitemShut {NoStop}%
\bibitem [{\citenamefont {Motrunich}\ \emph {et~al.}(2000)\citenamefont
  {Motrunich}, \citenamefont {Mau}, \citenamefont {Huse},\ and\ \citenamefont
  {Fisher}}]{motrunich-ising2d}%
  \BibitemOpen
  \bibfield  {author} {\bibinfo {author} {\bibfnamefont {O.}~\bibnamefont
  {Motrunich}}, \bibinfo {author} {\bibfnamefont {S.-C.}\ \bibnamefont {Mau}},
  \bibinfo {author} {\bibfnamefont {D.~A.}\ \bibnamefont {Huse}}, \ and\
  \bibinfo {author} {\bibfnamefont {D.~S.}\ \bibnamefont {Fisher}},\ }\href
  {\doibase 10.1103/PhysRevB.61.1160} {\bibfield  {journal} {\bibinfo
  {journal} {Phys. Rev. B}\ }\textbf {\bibinfo {volume} {61}},\ \bibinfo
  {pages} {1160} (\bibinfo {year} {2000})}\BibitemShut {NoStop}%
\bibitem [{\citenamefont {Hoyos}\ \emph
  {et~al.}(2007{\natexlab{a}})\citenamefont {Hoyos}, \citenamefont {Kotabage},\
  and\ \citenamefont {Vojta}}]{hoyos-kotabage-vojta-prl07}%
  \BibitemOpen
  \bibfield  {author} {\bibinfo {author} {\bibfnamefont {J.~A.}\ \bibnamefont
  {Hoyos}}, \bibinfo {author} {\bibfnamefont {C.}~\bibnamefont {Kotabage}}, \
  and\ \bibinfo {author} {\bibfnamefont {T.}~\bibnamefont {Vojta}},\ }\href
  {\doibase 10.1103/PhysRevLett.99.230601} {\bibfield  {journal} {\bibinfo
  {journal} {Phys. Rev. Lett.}\ }\textbf {\bibinfo {volume} {99}},\ \bibinfo
  {pages} {230601} (\bibinfo {year} {2007}{\natexlab{a}})}\BibitemShut
  {NoStop}%
\bibitem [{\citenamefont {Hooyberghs}\ \emph {et~al.}(2003)\citenamefont
  {Hooyberghs}, \citenamefont {Igl\'oi},\ and\ \citenamefont
  {Vanderzande}}]{hooyberghs-prl}%
  \BibitemOpen
  \bibfield  {author} {\bibinfo {author} {\bibfnamefont {J.}~\bibnamefont
  {Hooyberghs}}, \bibinfo {author} {\bibfnamefont {F.}~\bibnamefont {Igl\'oi}},
  \ and\ \bibinfo {author} {\bibfnamefont {C.}~\bibnamefont {Vanderzande}},\
  }\href {\doibase 10.1103/PhysRevLett.90.100601} {\bibfield  {journal}
  {\bibinfo  {journal} {Phys. Rev. Lett.}\ }\textbf {\bibinfo {volume} {90}},\
  \bibinfo {pages} {100601} (\bibinfo {year} {2003})}\BibitemShut {NoStop}%
\bibitem [{\citenamefont {Vojta}\ and\ \citenamefont
  {Hoyos}(2015)}]{vojta-hoyos-epl15}%
  \BibitemOpen
  \bibfield  {author} {\bibinfo {author} {\bibfnamefont {T.}~\bibnamefont
  {Vojta}}\ and\ \bibinfo {author} {\bibfnamefont {J.~A.}\ \bibnamefont
  {Hoyos}},\ }\href {http://stacks.iop.org/0295-5075/112/i=3/a=30002}
  {\bibfield  {journal} {\bibinfo  {journal} {EPL (Europhysics Letters)}\
  }\textbf {\bibinfo {volume} {112}},\ \bibinfo {pages} {30002} (\bibinfo
  {year} {2015})}\BibitemShut {NoStop}%
\bibitem [{\citenamefont {Monthus}(2017)}]{monthus-floquet}%
  \BibitemOpen
  \bibfield  {author} {\bibinfo {author} {\bibfnamefont {C.}~\bibnamefont
  {Monthus}},\ }\href {http://stacks.iop.org/1742-5468/2017/i=7/a=073301}
  {\bibfield  {journal} {\bibinfo  {journal} {Journal of Statistical Mechanics:
  Theory and Experiment}\ }\textbf {\bibinfo {volume} {2017}},\ \bibinfo
  {pages} {073301} (\bibinfo {year} {2017})}\BibitemShut {NoStop}%
\bibitem [{\citenamefont {Berdanier}\ \emph {et~al.}(2018)\citenamefont
  {Berdanier}, \citenamefont {Kolodrubetz}, \citenamefont {Parameswaran},\ and\
  \citenamefont {Vasseur}}]{berdanier-etal-pnas18}%
  \BibitemOpen
  \bibfield  {author} {\bibinfo {author} {\bibfnamefont {W.}~\bibnamefont
  {Berdanier}}, \bibinfo {author} {\bibfnamefont {M.}~\bibnamefont
  {Kolodrubetz}}, \bibinfo {author} {\bibfnamefont {S.~A.}\ \bibnamefont
  {Parameswaran}}, \ and\ \bibinfo {author} {\bibfnamefont {R.}~\bibnamefont
  {Vasseur}},\ }\href {\doibase 10.1073/pnas.1805796115} {\bibfield  {journal}
  {\bibinfo  {journal} {Proceedings of the National Academy of Sciences}\
  }\textbf {\bibinfo {volume} {115}},\ \bibinfo {pages} {9491} (\bibinfo {year}
  {2018})}\BibitemShut {NoStop}%
\bibitem [{\citenamefont {{S. K. Ma}}\ \emph {et~al.}(1979)\citenamefont {{S.
  K. Ma}}, \citenamefont {{C. Dasgupta}},\ and\ \citenamefont {{C. K.
  Hu}}}]{SDRG1}%
  \BibitemOpen
  \bibfield  {author} {\bibinfo {author} {\bibnamefont {{S. K. Ma}}}, \bibinfo
  {author} {\bibnamefont {{C. Dasgupta}}}, \ and\ \bibinfo {author}
  {\bibnamefont {{C. K. Hu}}},\ }\href {\doibase 10.1103/PhysRevLett.43.1434}
  {\bibfield  {journal} {\bibinfo  {journal} {Phys. Rev. Lett}\ }\textbf
  {\bibinfo {volume} {43}},\ \bibinfo {pages} {1434} (\bibinfo {year}
  {1979})}\BibitemShut {NoStop}%
\bibitem [{\citenamefont {Igl\'oi}\ and\ \citenamefont {{C.
  Monthus}}(2005)}]{SDRG2}%
  \BibitemOpen
  \bibfield  {author} {\bibinfo {author} {\bibfnamefont {F.}~\bibnamefont
  {Igl\'oi}}\ and\ \bibinfo {author} {\bibnamefont {{C. Monthus}}},\ }\href
  {\doibase 10.1016/j.physrep.2005.02.006} {\bibfield  {journal} {\bibinfo
  {journal} {Phys. Rep.}\ }\textbf {\bibinfo {volume} {412}},\ \bibinfo {pages}
  {277} (\bibinfo {year} {2005})}\BibitemShut {NoStop}%
\bibitem [{\citenamefont {{Igl{\'o}i}}\ and\ \citenamefont
  {{Monthus}}(2018)}]{igloi-monthus-review2-arxiv}%
  \BibitemOpen
  \bibfield  {author} {\bibinfo {author} {\bibfnamefont {F.}~\bibnamefont
  {{Igl{\'o}i}}}\ and\ \bibinfo {author} {\bibfnamefont {C.}~\bibnamefont
  {{Monthus}}},\ }\href@noop {} {\bibfield  {journal} {\bibinfo  {journal}
  {ArXiv e-prints}\ } (\bibinfo {year} {2018})},\ \Eprint
  {http://arxiv.org/abs/1806.07684} {arXiv:1806.07684 [cond-mat.dis-nn]}
  \BibitemShut {NoStop}%
\bibitem [{\citenamefont {Pich}\ \emph {et~al.}(1998)\citenamefont {Pich},
  \citenamefont {Young}, \citenamefont {Rieger},\ and\ \citenamefont
  {Kawashima}}]{pich-etal-prl98}%
  \BibitemOpen
  \bibfield  {author} {\bibinfo {author} {\bibfnamefont {C.}~\bibnamefont
  {Pich}}, \bibinfo {author} {\bibfnamefont {A.~P.}\ \bibnamefont {Young}},
  \bibinfo {author} {\bibfnamefont {H.}~\bibnamefont {Rieger}}, \ and\ \bibinfo
  {author} {\bibfnamefont {N.}~\bibnamefont {Kawashima}},\ }\href {\doibase
  10.1103/PhysRevLett.81.5916} {\bibfield  {journal} {\bibinfo  {journal}
  {Phys. Rev. Lett.}\ }\textbf {\bibinfo {volume} {81}},\ \bibinfo {pages}
  {5916} (\bibinfo {year} {1998})}\BibitemShut {NoStop}%
\bibitem [{\citenamefont {Shu}\ \emph {et~al.}(2016)\citenamefont {Shu},
  \citenamefont {Yao}, \citenamefont {Ke}, \citenamefont {Lin},\ and\
  \citenamefont {Sandvik}}]{shu-etal-prb16}%
  \BibitemOpen
  \bibfield  {author} {\bibinfo {author} {\bibfnamefont {Y.-R.}\ \bibnamefont
  {Shu}}, \bibinfo {author} {\bibfnamefont {D.-X.}\ \bibnamefont {Yao}},
  \bibinfo {author} {\bibfnamefont {C.-W.}\ \bibnamefont {Ke}}, \bibinfo
  {author} {\bibfnamefont {Y.-C.}\ \bibnamefont {Lin}}, \ and\ \bibinfo
  {author} {\bibfnamefont {A.~W.}\ \bibnamefont {Sandvik}},\ }\href {\doibase
  10.1103/PhysRevB.94.174442} {\bibfield  {journal} {\bibinfo  {journal} {Phys.
  Rev. B}\ }\textbf {\bibinfo {volume} {94}},\ \bibinfo {pages} {174442}
  (\bibinfo {year} {2016})}\BibitemShut {NoStop}%
\bibitem [{\citenamefont {Hamacher}\ \emph {et~al.}(2002)\citenamefont
  {Hamacher}, \citenamefont {Stolze},\ and\ \citenamefont {Wenzel}}]{hamacher}%
  \BibitemOpen
  \bibfield  {author} {\bibinfo {author} {\bibfnamefont {K.}~\bibnamefont
  {Hamacher}}, \bibinfo {author} {\bibfnamefont {J.}~\bibnamefont {Stolze}}, \
  and\ \bibinfo {author} {\bibfnamefont {W.}~\bibnamefont {Wenzel}},\ }\href
  {\doibase 10.1103/PhysRevLett.89.127202} {\bibfield  {journal} {\bibinfo
  {journal} {Phys. Rev. Lett.}\ }\textbf {\bibinfo {volume} {89}},\ \bibinfo
  {pages} {127202} (\bibinfo {year} {2002})}\BibitemShut {NoStop}%
\bibitem [{\citenamefont {Hida}(1996)}]{hidaale}%
  \BibitemOpen
  \bibfield  {author} {\bibinfo {author} {\bibfnamefont {K.}~\bibnamefont
  {Hida}},\ }\href {\doibase 10.1143/JPSJ.65.895} {\bibfield  {journal}
  {\bibinfo  {journal} {Journal of the Physical Society of Japan}\ }\textbf
  {\bibinfo {volume} {65}},\ \bibinfo {pages} {895} (\bibinfo {year}
  {1996})}\BibitemShut {NoStop}%
\bibitem [{\citenamefont {Ruggiero}\ \emph {et~al.}(2016)\citenamefont
  {Ruggiero}, \citenamefont {Alba},\ and\ \citenamefont
  {Calabrese}}]{ruggiero-etal-prb16}%
  \BibitemOpen
  \bibfield  {author} {\bibinfo {author} {\bibfnamefont {P.}~\bibnamefont
  {Ruggiero}}, \bibinfo {author} {\bibfnamefont {V.}~\bibnamefont {Alba}}, \
  and\ \bibinfo {author} {\bibfnamefont {P.}~\bibnamefont {Calabrese}},\ }\href
  {\doibase 10.1103/PhysRevB.94.035152} {\bibfield  {journal} {\bibinfo
  {journal} {Phys. Rev. B}\ }\textbf {\bibinfo {volume} {94}},\ \bibinfo
  {pages} {035152} (\bibinfo {year} {2016})}\BibitemShut {NoStop}%
\bibitem [{\citenamefont {Goldsborough}\ and\ \citenamefont
  {R\"omer}(2014)}]{rommerTNT}%
  \BibitemOpen
  \bibfield  {author} {\bibinfo {author} {\bibfnamefont {A.~M.}\ \bibnamefont
  {Goldsborough}}\ and\ \bibinfo {author} {\bibfnamefont {R.~A.}\ \bibnamefont
  {R\"omer}},\ }\href {\doibase 10.1103/PhysRevB.89.214203} {\bibfield
  {journal} {\bibinfo  {journal} {Phys. Rev. B}\ }\textbf {\bibinfo {volume}
  {89}},\ \bibinfo {pages} {214203} (\bibinfo {year} {2014})}\BibitemShut
  {NoStop}%
\bibitem [{\citenamefont {Goldsborough}\ and\ \citenamefont
  {Evenbly}(2017)}]{evenblyale}%
  \BibitemOpen
  \bibfield  {author} {\bibinfo {author} {\bibfnamefont {A.~M.}\ \bibnamefont
  {Goldsborough}}\ and\ \bibinfo {author} {\bibfnamefont {G.}~\bibnamefont
  {Evenbly}},\ }\href {\doibase 10.1103/PhysRevB.96.155136} {\bibfield
  {journal} {\bibinfo  {journal} {Phys. Rev. B}\ }\textbf {\bibinfo {volume}
  {96}},\ \bibinfo {pages} {155136} (\bibinfo {year} {2017})}\BibitemShut
  {NoStop}%
\bibitem [{\citenamefont {Doty}\ and\ \citenamefont
  {Fisher}(1992)}]{doty-fisher}%
  \BibitemOpen
  \bibfield  {author} {\bibinfo {author} {\bibfnamefont {C.~A.}\ \bibnamefont
  {Doty}}\ and\ \bibinfo {author} {\bibfnamefont {D.~S.}\ \bibnamefont
  {Fisher}},\ }\href {\doibase 10.1103/PhysRevB.45.2167} {\bibfield  {journal}
  {\bibinfo  {journal} {Phys. Rev. B}\ }\textbf {\bibinfo {volume} {45}},\
  \bibinfo {pages} {2167} (\bibinfo {year} {1992})}\BibitemShut {NoStop}%
\bibitem [{\citenamefont {{Quito}}\ \emph
  {et~al.}(2017{\natexlab{a}})\citenamefont {{Quito}}, \citenamefont {{Lopes}},
  \citenamefont {{Hoyos}},\ and\ \citenamefont {{Miranda}}}]{quito-etal-07}%
  \BibitemOpen
  \bibfield  {author} {\bibinfo {author} {\bibfnamefont {V.~L.}\ \bibnamefont
  {{Quito}}}, \bibinfo {author} {\bibfnamefont {P.~L.~S.}\ \bibnamefont
  {{Lopes}}}, \bibinfo {author} {\bibfnamefont {J.~A.}\ \bibnamefont
  {{Hoyos}}}, \ and\ \bibinfo {author} {\bibfnamefont {E.}~\bibnamefont
  {{Miranda}}},\ }\href@noop {} {\bibfield  {journal} {\bibinfo  {journal}
  {ArXiv e-prints}\ } (\bibinfo {year} {2017}{\natexlab{a}})},\ \Eprint
  {http://arxiv.org/abs/1711.04781} {arXiv:1711.04781 [cond-mat.str-el]}
  \BibitemShut {NoStop}%
\bibitem [{\citenamefont {{Quito}}\ \emph
  {et~al.}(2017{\natexlab{b}})\citenamefont {{Quito}}, \citenamefont {{Lopes}},
  \citenamefont {{Hoyos}},\ and\ \citenamefont {{Miranda}}}]{quito-etal-07b}%
  \BibitemOpen
  \bibfield  {author} {\bibinfo {author} {\bibfnamefont {V.~L.}\ \bibnamefont
  {{Quito}}}, \bibinfo {author} {\bibfnamefont {P.~L.~S.}\ \bibnamefont
  {{Lopes}}}, \bibinfo {author} {\bibfnamefont {J.~A.}\ \bibnamefont
  {{Hoyos}}}, \ and\ \bibinfo {author} {\bibfnamefont {E.}~\bibnamefont
  {{Miranda}}},\ }\href@noop {} {\bibfield  {journal} {\bibinfo  {journal}
  {ArXiv e-prints}\ } (\bibinfo {year} {2017}{\natexlab{b}})},\ \Eprint
  {http://arxiv.org/abs/1711.04783} {arXiv:1711.04783 [cond-mat.str-el]}
  \BibitemShut {NoStop}%
\bibitem [{\citenamefont {Lieb}\ \emph {et~al.}(1961)\citenamefont {Lieb},
  \citenamefont {Schultz},\ and\ \citenamefont {Mattis}}]{lieb-schultz-mattis}%
  \BibitemOpen
  \bibfield  {author} {\bibinfo {author} {\bibfnamefont {E.}~\bibnamefont
  {Lieb}}, \bibinfo {author} {\bibfnamefont {T.}~\bibnamefont {Schultz}}, \
  and\ \bibinfo {author} {\bibfnamefont {D.}~\bibnamefont {Mattis}},\ }\href
  {\doibase 10.1016/0003-4916(61)90115-4} {\bibfield  {journal} {\bibinfo
  {journal} {Ann. Phys.}\ }\textbf {\bibinfo {volume} {16}},\ \bibinfo {pages}
  {407} (\bibinfo {year} {1961})}\BibitemShut {NoStop}%
\bibitem [{\citenamefont {Calabrese}\ and\ \citenamefont
  {Cardy}(2004)}]{cardyentan}%
  \BibitemOpen
  \bibfield  {author} {\bibinfo {author} {\bibfnamefont {P.}~\bibnamefont
  {Calabrese}}\ and\ \bibinfo {author} {\bibfnamefont {J.}~\bibnamefont
  {Cardy}},\ }\href {http://stacks.iop.org/1742-5468/2004/i=06/a=P06002}
  {\bibfield  {journal} {\bibinfo  {journal} {Journal of Statistical Mechanics:
  Theory and Experiment}\ }\textbf {\bibinfo {volume} {2004}},\ \bibinfo
  {pages} {P06002} (\bibinfo {year} {2004})}\BibitemShut {NoStop}%
\bibitem [{\citenamefont {Vidal}\ \emph {et~al.}(2003)\citenamefont {Vidal},
  \citenamefont {Latorre}, \citenamefont {Rico},\ and\ \citenamefont
  {Kitaev}}]{cvidal}%
  \BibitemOpen
  \bibfield  {author} {\bibinfo {author} {\bibfnamefont {G.}~\bibnamefont
  {Vidal}}, \bibinfo {author} {\bibfnamefont {J.~I.}\ \bibnamefont {Latorre}},
  \bibinfo {author} {\bibfnamefont {E.}~\bibnamefont {Rico}}, \ and\ \bibinfo
  {author} {\bibfnamefont {A.}~\bibnamefont {Kitaev}},\ }\href {\doibase
  10.1103/PhysRevLett.90.227902} {\bibfield  {journal} {\bibinfo  {journal}
  {Phys. Rev. Lett.}\ }\textbf {\bibinfo {volume} {90}},\ \bibinfo {pages}
  {227902} (\bibinfo {year} {2003})}\BibitemShut {NoStop}%
\bibitem [{\citenamefont {Korepin}(2004)}]{prlkorepin}%
  \BibitemOpen
  \bibfield  {author} {\bibinfo {author} {\bibfnamefont {V.~E.}\ \bibnamefont
  {Korepin}},\ }\href {\doibase 10.1103/PhysRevLett.92.096402} {\bibfield
  {journal} {\bibinfo  {journal} {Phys. Rev. Lett.}\ }\textbf {\bibinfo
  {volume} {92}},\ \bibinfo {pages} {096402} (\bibinfo {year}
  {2004})}\BibitemShut {NoStop}%
\bibitem [{\citenamefont {Amico}\ \emph {et~al.}(2008)\citenamefont {Amico},
  \citenamefont {Fazio}, \citenamefont {Osterloh},\ and\ \citenamefont
  {Vedral}}]{revfazio}%
  \BibitemOpen
  \bibfield  {author} {\bibinfo {author} {\bibfnamefont {L.}~\bibnamefont
  {Amico}}, \bibinfo {author} {\bibfnamefont {R.}~\bibnamefont {Fazio}},
  \bibinfo {author} {\bibfnamefont {A.}~\bibnamefont {Osterloh}}, \ and\
  \bibinfo {author} {\bibfnamefont {V.}~\bibnamefont {Vedral}},\ }\href
  {\doibase 10.1103/RevModPhys.80.517} {\bibfield  {journal} {\bibinfo
  {journal} {Rev. Mod. Phys.}\ }\textbf {\bibinfo {volume} {80}},\ \bibinfo
  {pages} {517} (\bibinfo {year} {2008})}\BibitemShut {NoStop}%
\bibitem [{\citenamefont {Eisert}\ \emph {et~al.}(2010)\citenamefont {Eisert},
  \citenamefont {Cramer},\ and\ \citenamefont {Plenio}}]{RMP82-277}%
  \BibitemOpen
  \bibfield  {author} {\bibinfo {author} {\bibfnamefont {J.}~\bibnamefont
  {Eisert}}, \bibinfo {author} {\bibfnamefont {M.}~\bibnamefont {Cramer}}, \
  and\ \bibinfo {author} {\bibfnamefont {M.~B.}\ \bibnamefont {Plenio}},\
  }\href {\doibase 10.1103/RevModPhys.82.277} {\bibfield  {journal} {\bibinfo
  {journal} {Rev. Mod. Phys.}\ }\textbf {\bibinfo {volume} {82}},\ \bibinfo
  {pages} {277} (\bibinfo {year} {2010})}\BibitemShut {NoStop}%
\bibitem [{\citenamefont {Calabrese}\ and\ \citenamefont
  {Cardy}(2009)}]{entroreviewcalabrese}%
  \BibitemOpen
  \bibfield  {author} {\bibinfo {author} {\bibfnamefont {P.}~\bibnamefont
  {Calabrese}}\ and\ \bibinfo {author} {\bibfnamefont {J.}~\bibnamefont
  {Cardy}},\ }\href {http://stacks.iop.org/1751-8121/42/i=50/a=504005}
  {\bibfield  {journal} {\bibinfo  {journal} {Journal of Physics A:
  Mathematical and Theoretical}\ }\textbf {\bibinfo {volume} {42}},\ \bibinfo
  {pages} {504005} (\bibinfo {year} {2009})}\BibitemShut {NoStop}%
\bibitem [{\citenamefont {{G. Refael}}\ and\ \citenamefont {{J. E.
  Moore}}(2004)}]{Mooreale}%
  \BibitemOpen
  \bibfield  {author} {\bibinfo {author} {\bibnamefont {{G. Refael}}}\ and\
  \bibinfo {author} {\bibnamefont {{J. E. Moore}}},\ }\href {\doibase
  10.1103/PhysRevB.76.024419} {\bibfield  {journal} {\bibinfo  {journal} {Phys.
  Rev. Lett.}\ }\textbf {\bibinfo {volume} {93}},\ \bibinfo {pages} {260602}
  (\bibinfo {year} {2004})}\BibitemShut {NoStop}%
\bibitem [{\citenamefont {Laflorencie}(2005)}]{laflorencie-entanglement}%
  \BibitemOpen
  \bibfield  {author} {\bibinfo {author} {\bibfnamefont {N.}~\bibnamefont
  {Laflorencie}},\ }\href {\doibase 10.1103/PhysRevB.72.140408} {\bibfield
  {journal} {\bibinfo  {journal} {Phys. Rev. B}\ }\textbf {\bibinfo {volume}
  {72}},\ \bibinfo {pages} {140408} (\bibinfo {year} {2005})}\BibitemShut
  {NoStop}%
\bibitem [{\citenamefont {Hoyos}\ \emph
  {et~al.}(2007{\natexlab{b}})\citenamefont {Hoyos}, \citenamefont {Vieira},
  \citenamefont {Laflorencie},\ and\ \citenamefont {Miranda}}]{abel-ale}%
  \BibitemOpen
  \bibfield  {author} {\bibinfo {author} {\bibfnamefont {J.~A.}\ \bibnamefont
  {Hoyos}}, \bibinfo {author} {\bibfnamefont {A.~P.}\ \bibnamefont {Vieira}},
  \bibinfo {author} {\bibfnamefont {N.}~\bibnamefont {Laflorencie}}, \ and\
  \bibinfo {author} {\bibfnamefont {E.}~\bibnamefont {Miranda}},\ }\href
  {\doibase 10.1103/PhysRevB.76.174425} {\bibfield  {journal} {\bibinfo
  {journal} {Phys. Rev. B}\ }\textbf {\bibinfo {volume} {76}},\ \bibinfo
  {pages} {174425} (\bibinfo {year} {2007}{\natexlab{b}})}\BibitemShut
  {NoStop}%
\bibitem [{\citenamefont {{M. Fagotti}}\ \emph {et~al.}(2011)\citenamefont {{M.
  Fagotti}}, \citenamefont {{P. Calabrese}},\ and\ \citenamefont {{J. E.
  Moore}}}]{calabreserandom}%
  \BibitemOpen
  \bibfield  {author} {\bibinfo {author} {\bibnamefont {{M. Fagotti}}},
  \bibinfo {author} {\bibnamefont {{P. Calabrese}}}, \ and\ \bibinfo {author}
  {\bibnamefont {{J. E. Moore}}},\ }\href {\doibase 10.1103/PhysRevB.83.045110}
  {\bibfield  {journal} {\bibinfo  {journal} {Phys. Rev. B}\ }\textbf {\bibinfo
  {volume} {83}},\ \bibinfo {pages} {045110} (\bibinfo {year}
  {2011})}\BibitemShut {NoStop}%
\bibitem [{\citenamefont {Quito}\ \emph {et~al.}(2015)\citenamefont {Quito},
  \citenamefont {Hoyos},\ and\ \citenamefont
  {Miranda}}]{quito-hoyos-miranda-prl15}%
  \BibitemOpen
  \bibfield  {author} {\bibinfo {author} {\bibfnamefont {V.~L.}\ \bibnamefont
  {Quito}}, \bibinfo {author} {\bibfnamefont {J.~A.}\ \bibnamefont {Hoyos}}, \
  and\ \bibinfo {author} {\bibfnamefont {E.}~\bibnamefont {Miranda}},\ }\href
  {\doibase 10.1103/PhysRevLett.115.167201} {\bibfield  {journal} {\bibinfo
  {journal} {Phys. Rev. Lett.}\ }\textbf {\bibinfo {volume} {115}},\ \bibinfo
  {pages} {167201} (\bibinfo {year} {2015})}\BibitemShut {NoStop}%
\bibitem [{\citenamefont {{J. A. Hoyos}}\ and\ \citenamefont {{E.
  Miranda}}(2004)}]{netoSUNdesor}%
  \BibitemOpen
  \bibfield  {author} {\bibinfo {author} {\bibnamefont {{J. A. Hoyos}}}\ and\
  \bibinfo {author} {\bibnamefont {{E. Miranda}}},\ }\href {\doibase
  10.1103/PhysRevB.70.180401} {\bibfield  {journal} {\bibinfo  {journal} {Phys.
  Rev. B}\ }\textbf {\bibinfo {volume} {70}},\ \bibinfo {pages} {180401(R)}
  (\bibinfo {year} {2004})}\BibitemShut {NoStop}%
\bibitem [{\citenamefont {Peschel}(2003)}]{peschel-jpa03}%
  \BibitemOpen
  \bibfield  {author} {\bibinfo {author} {\bibfnamefont {I.}~\bibnamefont
  {Peschel}},\ }\href {http://stacks.iop.org/0305-4470/36/i=14/a=101}
  {\bibfield  {journal} {\bibinfo  {journal} {J. Phys. A: Math. Gen.}\ }\textbf
  {\bibinfo {volume} {36}},\ \bibinfo {pages} {L205} (\bibinfo {year}
  {2003})}\BibitemShut {NoStop}%
\bibitem [{\citenamefont {Hoyos}\ and\ \citenamefont
  {Rigolin}(2006)}]{hoyos-rigolin}%
  \BibitemOpen
  \bibfield  {author} {\bibinfo {author} {\bibfnamefont {J.~A.}\ \bibnamefont
  {Hoyos}}\ and\ \bibinfo {author} {\bibfnamefont {G.}~\bibnamefont
  {Rigolin}},\ }\href {\doibase http://dx.doi.org/10.1103/PhysRevA.74.062324}
  {\bibfield  {journal} {\bibinfo  {journal} {Phys. Rev. A}\ }\textbf {\bibinfo
  {volume} {74}},\ \bibinfo {pages} {062324} (\bibinfo {year}
  {2006})}\BibitemShut {NoStop}%
\bibitem [{\citenamefont {{Getelina}}\ \emph {et~al.}(2018)\citenamefont
  {{Getelina}}, \citenamefont {{de Oliveira}},\ and\ \citenamefont
  {{Hoyos}}}]{getelina-etal-2017}%
  \BibitemOpen
  \bibfield  {author} {\bibinfo {author} {\bibfnamefont {J.~C.}\ \bibnamefont
  {{Getelina}}}, \bibinfo {author} {\bibfnamefont {T.~R.}\ \bibnamefont {{de
  Oliveira}}}, \ and\ \bibinfo {author} {\bibfnamefont {J.~A.}\ \bibnamefont
  {{Hoyos}}},\ }\href {\doibase https://doi.org/10.1016/j.physleta.2018.08.003}
  {\bibfield  {journal} {\bibinfo  {journal} {Physics Letters A}\ }\textbf
  {\bibinfo {volume} {382}},\ \bibinfo {pages} {2799} (\bibinfo {year}
  {2018})}\BibitemShut {NoStop}%
\bibitem [{Note1()}]{Note1}%
  \BibitemOpen
  \bibinfo {note} {A similar situation may also happen in frustrated systems,
  where there are several states with energies very close to the ground
  energy.}\BibitemShut {Stop}%
\bibitem [{\citenamefont {{S. R. White}}(1996)}]{dmrg2dwhite}%
  \BibitemOpen
  \bibfield  {author} {\bibinfo {author} {\bibnamefont {{S. R. White}}},\
  }\href {\doibase 10.1103/PhysRevLett.77.3633} {\bibfield  {journal} {\bibinfo
   {journal} {Phys. Rev. Lett.}\ }\textbf {\bibinfo {volume} {77}},\ \bibinfo
  {pages} {3633} (\bibinfo {year} {1996})}\BibitemShut {NoStop}%
\bibitem [{\citenamefont {White}\ and\ \citenamefont
  {Feiguin}(2004)}]{tDMRG-white}%
  \BibitemOpen
  \bibfield  {author} {\bibinfo {author} {\bibfnamefont {S.~R.}\ \bibnamefont
  {White}}\ and\ \bibinfo {author} {\bibfnamefont {A.~E.}\ \bibnamefont
  {Feiguin}},\ }\href {\doibase 10.1103/PhysRevLett.93.076401} {\bibfield
  {journal} {\bibinfo  {journal} {Phys. Rev. Lett.}\ }\textbf {\bibinfo
  {volume} {93}},\ \bibinfo {pages} {076401} (\bibinfo {year}
  {2004})}\BibitemShut {NoStop}%
\bibitem [{\citenamefont {{P. Henelius}}\ and\ \citenamefont {{S. M.
  Girvin}}(1998)}]{girvenale}%
  \BibitemOpen
  \bibfield  {author} {\bibinfo {author} {\bibnamefont {{P. Henelius}}}\ and\
  \bibinfo {author} {\bibnamefont {{S. M. Girvin}}},\ }\href {\doibase
  10.1103/PhysRevB.57.11457} {\bibfield  {journal} {\bibinfo  {journal} {Phys.
  Rev. B}\ }\textbf {\bibinfo {volume} {57}},\ \bibinfo {pages} {11457}
  (\bibinfo {year} {1998})}\BibitemShut {NoStop}%
\bibitem [{\citenamefont {{N. Laflorencie}}\ \emph {et~al.}(2004)\citenamefont
  {{N. Laflorencie}}, \citenamefont {{H. Rieger}}, \citenamefont {{A. W.
  Sandvik}},\ and\ \citenamefont {{P. Henelius}}}]{laflorencie-ale}%
  \BibitemOpen
  \bibfield  {author} {\bibinfo {author} {\bibnamefont {{N. Laflorencie}}},
  \bibinfo {author} {\bibnamefont {{H. Rieger}}}, \bibinfo {author}
  {\bibnamefont {{A. W. Sandvik}}}, \ and\ \bibinfo {author} {\bibnamefont {{P.
  Henelius}}},\ }\href {\doibase 10.1103/PhysRevB.70.054430} {\bibfield
  {journal} {\bibinfo  {journal} {Phys. Rev. B}\ ,\ \bibinfo {pages} {054430}}
  (\bibinfo {year} {2004})}\BibitemShut {NoStop}%
\bibitem [{\citenamefont {{A. Luther}}\ and\ \citenamefont {{I.
  Peschel}}(1975)}]{lutherpeschel2}%
  \BibitemOpen
  \bibfield  {author} {\bibinfo {author} {\bibnamefont {{A. Luther}}}\ and\
  \bibinfo {author} {\bibnamefont {{I. Peschel}}},\ }\href {\doibase
  10.1103/PhysRevB.12.3908} {\bibfield  {journal} {\bibinfo  {journal} {Phys.
  Rev. B}\ }\textbf {\bibinfo {volume} {12}},\ \bibinfo {pages} {3908}
  (\bibinfo {year} {1975})}\BibitemShut {NoStop}%
\bibitem [{\citenamefont {{C. Itoi}}\ and\ \citenamefont {{M.-H.
  Kato}}(1997)}]{itoi-biqua-clean}%
  \BibitemOpen
  \bibfield  {author} {\bibinfo {author} {\bibnamefont {{C. Itoi}}}\ and\
  \bibinfo {author} {\bibnamefont {{M.-H. Kato}}},\ }\href {\doibase
  10.1103/PhysRevB.55.8295} {\bibfield  {journal} {\bibinfo  {journal} {Phys.
  Rev. B}\ }\textbf {\bibinfo {volume} {55}},\ \bibinfo {pages} {8295}
  (\bibinfo {year} {1997})}\BibitemShut {NoStop}%
\end{thebibliography}%

\end{document}